\documentclass{article}
\usepackage{arxiv}
\usepackage{multirow}
\usepackage{tabularx}
\usepackage{makecell}
\usepackage[utf8]{inputenc}
\usepackage[T1]{fontenc}
\usepackage{hyperref}
\usepackage{url}
\usepackage{booktabs}
\usepackage{amsfonts}
\usepackage{amssymb}
\usepackage{nicefrac}
\usepackage{microtype}
\usepackage{graphicx}
\usepackage{natbib}
\usepackage{doi}
\usepackage[fleqn]{amsmath}
\usepackage{tikz}
\usepackage[table]{xcolor}
\usetikzlibrary{arrows.meta, positioning}
\definecolor{Exo}{HTML}{FFEFD3}
\definecolor{Med}{HTML}{CBF5EA}
\definecolor{Beh}{HTML}{CFDEF5}
\definecolor{Outc}{HTML}{FFE4D3}
\definecolor{Bias}{HTML}{97745F}

\title{Space-time accessibility supports participation in after-work leisure activities}

\author{ \href{https://orcid.org/0000-0002-6982-1654}{\includegraphics[scale=0.06]{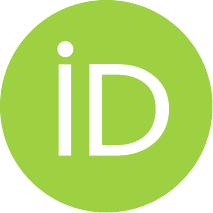}\hspace{1mm}Yuan~Liao}\thanks{Corresponding author. Also affiliated with Department of Applied Mathematics and Computer Science, Technical University of Denmark, Lyngby, Denmark.} \\
	Department of Space, Earth and Environment\\
	Chalmers University of Technology\\
	Gothenburg, Sweden \\
	\texttt{yuan.liao@chalmers.se} \\
    \And
	\href{https://orcid.org/0000-0003-2125-7465}{\includegraphics[scale=0.06]{orcid.pdf}\hspace{1mm}Rafael~H. M.~Pereira} \\
	Institute for Applied Economic Research (Ipea) - Brazil\\
	Data Science Division, Brazil\\
	\texttt{rafael.pereira@ipea.gov.br} \\
	\And
	\href{https://orcid.org/0000-0001-6671-2578}{\includegraphics[scale=0.06]{orcid.pdf}\hspace{1mm}Jorge~Gil} \\
	Department of Architecture and Civil Engineering\\
	Chalmers University of Technology\\
	Gothenburg, Sweden\\
	\texttt{jorge.gil@chalmers.se} \\
    \And
	\href{https://orcid.org/0000-0002-6719-0764}{\includegraphics[scale=0.06]{orcid.pdf}\hspace{1mm}Silvia~De~Sojo~Caso} \\
	Department of Applied Mathematics and Computer Science\\
	Technical University of Denmark\\
    Lyngby, Denmark\\
	\texttt{sdsc@dtu.dk} \\
    \And
	\href{https://orcid.org/0000-0001-6003-1165}{\includegraphics[scale=0.06]{orcid.pdf}\hspace{1mm}Laura~Alessandretti} \\
	Department of Applied Mathematics and Computer Science\\
	Technical University of Denmark\\
    Lyngby, Denmark\\
	\texttt{lauale@dtu.dk} \\
}

\renewcommand{\shorttitle}{Space-time accessibility supports participation in after-work leisure activities}

\hypersetup{
pdftitle={Space-time accessibility supports participation in after-work leisure activities},
pdfsubject={transport accessibility, structural equation modeling},
pdfauthor={Yuan Liao, Rafael H. M. Pereira, Jorge Gil, Silvia De Sojo Caso, Laura Alessandretti},
pdfkeywords={space-time accessibility, third-place activities, structural equation modeling, transportation equity},
}

\begin{document}
\newcommand{\tabincell}[2]{
\begin{tabular}{@{}#1@{}}#2\end{tabular}
}
\maketitle

\begin{abstract}
Understanding how accessibility shapes participation in leisure activities is central to promoting inclusive and vibrant urban life.
Conventional accessibility measures often focus on potential access from fixed home locations, overlooking the constraints and opportunities embedded in daily routines.
In this study, we apply a space–time accessibility (STA) metric rooted in the capability approach, capturing feasible leisure opportunities between home and work given a certain time budget, individual transport modes, and urban infrastructure.
Using high-resolution GPS data from 2,415 working residents in the Paris region, we assess how STA influences leisure participation during weekdays, measured as the diversity of leisure locations visited and activity duration.
Observed destination choices confirm that most individuals select leisure locations within their STA-defined opportunity sets, validating the metric as a proxy for capability sets.
Structural equation modeling shows that STA exerts a significant positive total effect on leisure participation ($\beta = 0.14$, $p < .001$), driven by a significant direct effect ($\beta = 0.18$, $p < .001$) that is only modestly offset by an indirect pathway through reduced travel time ($\beta = -0.04$, $p < .01$).
Individual attributes also directly shape participation: active mode use and higher education promote leisure engagement, while local poverty and caregiving responsibilities constrain it.
These findings highlight the value of person-centered, capability-informed accessibility metrics for understanding inequalities in urban mobility and informing transport planning strategies that expand real freedoms to participate in social life across diverse population groups.
\end{abstract}


\section{Introduction}\label{sec:introduction}
Transport accessibility is central to inclusive cities through its influence on individual mobility and activity participation \citep{allen2020planning,luz2022does, liao2025socio}.
By shaping how easily people can reach different locations and activities, spatial accessibility defines the scope of places and social environments individuals can access \citep{pereira2017distributive}, particularly activities taking place in third places, that is, informal settings outside home and work (e.g., cafés, community centres, parks) that foster meaningful social interactions \citep{oldenburgThirdPlace1982}.
When transport opportunities are limited, the risk of social exclusion rises for certain groups \citep{luz2022understanding, gallego2023social}.
Studies have shown that greater proximity to certain destinations is associated with higher participation rates in those activities \citep[e.g.,][]{reimers2014proximity}.
Yet, it remains unclear whether access to transportation effectively is associated with more participation in third-place activities, and how individual capabilities and time constraints shape this relationship. \par

Although third-place activities are crucial for social inclusion and well-being, they have received less scholarly attention than commuting and essential travel domains in accessibility literature \citep{luz2022does,tomasiello2023unfolding}.
Visits to third places affect individuals' social exposure by linking them to wider society beyond institutional settings \citep{oldenburgThirdPlace1982}.
Such exposure often depends on people's flexibility in allocating time to discretionary activities---that is, non-obligatory activities such as socializing, leisure, or informal errands that they choose to engage in.
Recent studies reflect the growing attention to third-place activities in accessibility research, e.g., evaluating inequalities in access to destinations like parks \citep{tomasiello2023unfolding,barboza2024comparative}.
Among the third-place activities, this paper focuses specifically on leisure activities.
This choice highlights that transport systems must accommodate the diverse and time-sensitive mobility demands associated with activities that improve individuals' well-being \citep{lee2022third}. \par

Accessibility to leisure opportunities is shaped by transport systems, which play a key role in enabling activity participation, yet it is usually examined in a static manner \citep{luz2022understanding}.
Home- or work-based accessibility measures, typically operationalized through network-based models of potential access \citep{levinson2020transport}, estimate how many leisure destinations can be reached within a given travel-time threshold.
Most studies in the literature utilize such home-based and static measures of transport accessibility \citep{ryan2023accessibility}.
There are two major limitations of static measures.
First, this stream largely overlooks trip chaining and trips that originate from non-home locations, particularly workplaces.
Many leisure activities, central to urban vibrancy and the fostering of social interactions \citep{botta2021modelling}, take place after work rather than from home \citep{mcguckin1999examining, farber2013social, barboza2024comparative}.
Second, static approaches neglect the dynamics of individuals being in motion throughout their daily activities.
One of the key factors shaping individuals' ability to engage in leisure activities is their travel time budget, defined as the amount of time people are willing to spend traveling each day.
Individuals who pursue a wider range of activities---particularly discretionary ones, such as leisure, social engagements, or personal errands---tend to accumulate longer daily travel time budgets, as these additional and geographically dispersed activities extend the time spent traveling \citep{joly2016intensive}.
Conversely, individuals with tighter activity schedules or fewer discretionary commitments exhibit more limited daily travel-time budgets and remain anchored to a narrower set of proximate destinations.
Activity locations in proximity reduce the required travel effort for reaching comparable opportunities, thereby lowering the effective time budget needed to satisfy activity demand \citep{chen2021effects}. \par

Unlike home- or work-based measures, space-time accessibility metrics account for the spatio-temporal constraints of an individual's time–space prism \citep{kim2003space, victoriano2020time, saraiva2022accessibility}, offering a more realistic proxy for their ability to access leisure opportunities.
However, space–time accessibility metrics are data-intensive and vary substantially across individuals, requiring detailed information to compute, including home and work locations \citep{saraiva2022accessibility}.
The resulting measures also require careful interpretation, as understanding them often depends on additional personal attributes to disentangle the complex factors shaping accessibility outcomes \citep{chen2021effects}.
Consequently, applying these metrics---and studying their downstream effects on activity participation---generally requires high-resolution datasets, such as the one used in the present study. \par

\textit{To date, our understanding of how transport accessibility shapes leisure activity participation remains largely static.}
This study draws on a unique geolocation dataset capturing multi-day movement trajectories for over 2,000 residents of the Greater Paris region, alongside detailed socio-demographic and household information.
In this study, we i) quantify individual space–time accessibility to leisure opportunities, ii) examine its explanatory power for total travel time and leisure activity participation (location diversity), and iii) assess how these relationships vary across population groups. \par

The remainder of this manuscript is organized as follows.
The introduction continues with a brief review of related work (Section \ref{sec:related_work}), followed by an explanation of the applied conceptual framework (Section \ref{sec:concept}).
We then describe the materials (Section \ref{sec:materials}) and methodology (Section \ref{sec:methods}), followed by Section \ref{sec:results}, which presents the results and accompanying discussion.
The paper concludes with a summary of the main findings in the Conclusion section.

\subsection{Related work}\label{sec:related_work}
Third-place activities play a central role in urban life, yet their use and relevance are strongly structured by individuals' home and work places and personal circumstances.
Third-place activities---such as gym, café, and park---significantly shape social interactions, identity, and community connectedness \citep{oldenburgThirdPlace1982, farber2013social}.
For example, some third places, such as shopping centres, strategically located to connect diverse neighbourhoods, could attract visitors across socioeconomic groups and foster inter-group exposure \citep{nilforoshan2023human}.
Institutional locations, i.e., homes and workplaces, serve as anchors that structure daily mobility patterns, constraining the accessibility and relevance of third places \citep{miller1991modelling}.
The effect of these anchors on mobility is intertwined with individual attributes, lifestyle, and activity demand \citep{luz2022understanding, barboza2024comparative}. \par

Leisure activities, which typically happen in third places, differ behaviorally from the mandatory activities that anchor daily schedules.
In activity-based frameworks, home and work function as fixed anchors: spatially and temporally rigid commitments around which the rest of the daily program is organized \citep{schwanen2008fixed}.
Leisure activities, by contrast, are characterized by high scheduling flexibility, spatial substitutability, and optionality: they can be more easily relocated, rescheduled, shortened, or dropped altogether when time pressure mounts \citep{ettema2007modelling,schwanen2008fixed}.
These properties mean that leisure participation is especially sensitive to residual time budgets, the discretionary time that remains after mandatory commitments and commuting have been satisfied \citep{kitamura1984model,joly2016intensive}.
Unlike commuting trips, where the destination is fixed and the decision reduces to route and mode, leisure trips involve simultaneous choices of whether, when, where, and for how long to participate, making them inherently more responsive to variations in accessibility \citep{ettema2007modelling}. \par

A growing body of evidence indicates a strong correlation between transportation accessibility and physical activity participation.
Higher accessibility levels are strongly correlated with participation in total, mandatory, and discretionary activities \citep{fransen2018spatio, luz2022does}.
Proximity to walkable areas promotes higher levels of physical activity \citep{althoff2025countrywide}.
Improving accessibility in neighborhoods with concentrations of low-income or carless households located outside major transit corridors has also been shown to increase daily activity participation \citep{allen2020planning}.
The behavioral distinctiveness of leisure from the other mandatory activities makes leisure a revealing lens for evaluating how well transport systems convert potential access into actual participation: rigid activities will be undertaken regardless of accessibility conditions, whereas flexible activities amplify the signal of accessibility differences across individuals and places.
Indeed, studies have found that enhancing public transit accessibility can particularly benefit participation in discretionary activities among disadvantaged groups \citep[e.g.,][]{zhang2022eliminating}. \par

Nevertheless, most existing studies rely on home-based cumulative opportunity \citep{allen2020planning,luz2022does} or simple proximity measures \cite{mccormack2011search} of accessibility, treating individual attributes separately.
This approach is essentially static and overlooks the mechanisms through which accessibility shapes mobility.
For example, while greater job accessibility may be associated with higher leisure participation at the aggregate level \citep{luz2022does}, such a misalignment between the activity types used to measure accessibility and those used to measure participation obscures policy relevance and undermines individual-level insights. \par

In contrast, space–time accessibility is a person-based approach that enables the examination of intra-group and intra-location heterogeneities, thereby offering a deeper understanding of accessibility inequalities across individuals \citep{neutens2010equity, saraiva2022accessibility}.
It builds on two core concepts: an individual's trajectory in space and time (space-time path), and the set of potential trajectories an individual could take given temporal and spatial constraints (anchors) \citep{kwan1998space}.
They define the geographical extent of accessible opportunities (space-time prisms).
Measures of space–time accessibility range from direct depictions of prisms \citep{miller1991modelling} to empirically scalable methods based on travel time budgets \citep{saraiva2022accessibility}.
Recent advances in space–time accessibility account for the feasibility of chaining activities between home and work, given an individual's travel time budget, residential and workplace locations, and the built environment \citep{saraiva2022accessibility, barboza2024comparative}.
This formulation reflects realistic mobility patterns, such as engaging in leisure activities on the way home from work, and thus provides a useful proxy for assessing both mobility efficiency and opportunities for third-place participation. \par

To disentangle the complex pathways through which space–time accessibility shapes leisure participation, we employ Structural Equation Modeling (SEM), a widely used approach in travel behavior research \citep{Golob_2003, Kroesen_Van_Wee_2022, Song_et_al_2016}.
SEM is particularly adept at dissecting intricate causal structures \citep{Golob_2003, Kroesen_Van_Wee_2022}.
For instance, SEM has been applied to explore the interdependencies among land use, socio-demographic attributes, and travel patterns \citep{Song_et_al_2016}, and to model mediating variables such as car ownership when analyzing the built environment's influence on travel behavior \citep{Zhang_et_al_2025}.
This method has been shown to provide valuable insights into how built environment characteristics---including density, diversity, connectivity, and accessibility---affect various outcomes, such as individual transport emissions, health outcomes, and rural income, by shaping residents' travel choices and capabilities \citep{Azmoodeh_et_al_2023, Kroesen_Van_Wee_2022, Li_et_al_2024, Song_et_al_2016}.

\section{Conceptual framework}\label{sec:concept}
\begin{figure}[!ht]
\centering
\includegraphics[width=0.7\linewidth]{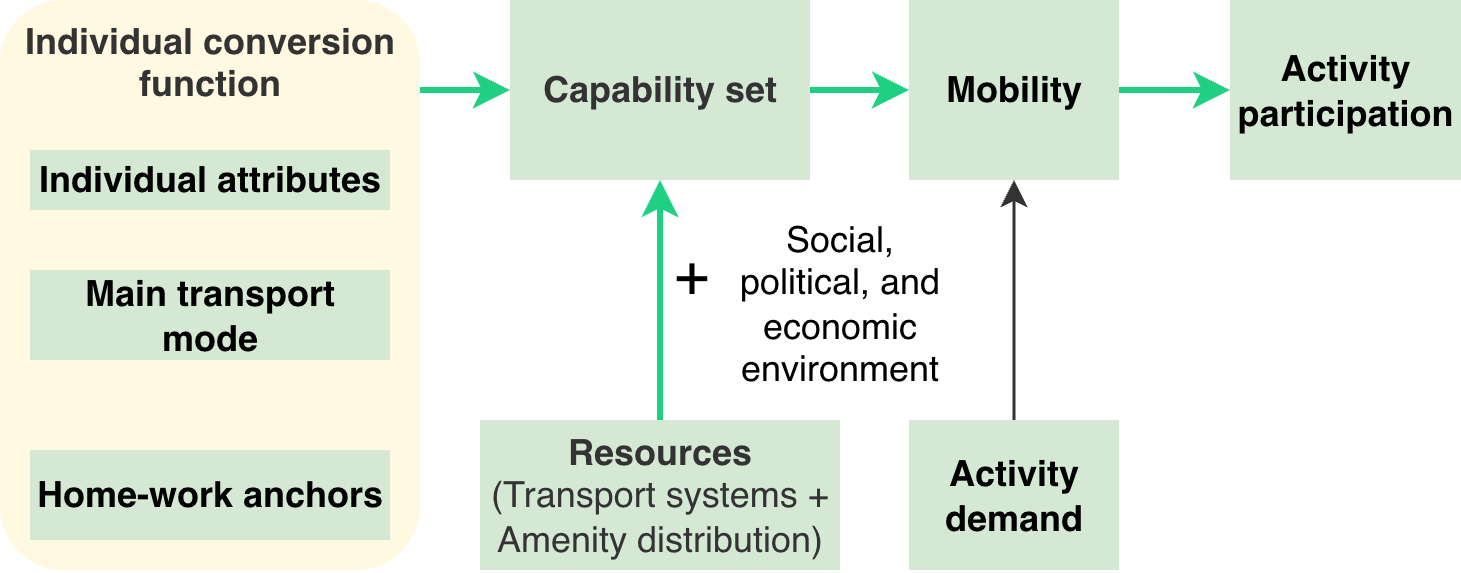}
\caption{\textbf{Accessibility as a human capability.}
Conceptual framework illustrating how we conceptualize space–time accessibility (STA) as a human capability, adapted from \cite{luz2022understanding,luz2022does}.
The diagram shows how factors influencing activity participation interact with capability sets.}
\label{fig:concept}
\end{figure}

In this study, we operationalize the notion of accessibility as a human capability using the space–time accessibility (STA) framework \citep{pereira2017distributive, luz2022understanding}.
In this approach, individuals' time budgets and mobility resources (e.g., main transport mode, amenity distribution) interact to shape the opportunity sets that one can reach, including third-place activities such as leisure (Figure \ref{fig:concept}). \par

First, individuals' conversion functions, shaped by personal attributes and constraints such as household responsibilities, education, and mobility skills, determine the inputs to the capability set.
Specifically, sociodemographic characteristics shape the two building blocks of space–time accessibility: residential and workplace locations (space-time anchors) and the main transport mode (the mobility resource).
The spatial-temporal dimension of capability --- that is, the leisure opportunities reachable within a given time budget, mode, and set of anchors --- thus emerges from these conversion processes.
Together, the conversion function and resources (transport systems, amenity distributions) produce the capability set: the set of leisure destinations one could feasibly visit on a given trip chain.
Second, this potential is realized through actual travel decisions, in which activity demand and time allocation jointly determine which opportunities are pursued.
Third, mobility enables participation in activities at the destinations reached.
This resource → potential → realization → participation logic motivates treating the capability set and mobility as distinct constructs: the capability set captures what is spatially and temporally feasible, while mobility captures the behavioral process through which feasible opportunities are selectively converted into actual engagement. \par

A key insight here is that these two processes (accessibility and mobility) can have opposing effects on leisure diversity.
A larger capability set expands options (positive effect), but individuals in high-accessibility settings may also travel less (because nearby opportunities suffice), and lower travel time is associated with lower diversity.
This countervailing mechanism, which we demonstrate empirically through SEM decomposition, is the core advancement over \cite{luz2022understanding}, who examined the link between mobility, activity participation, and wellbeing without quantitatively disentangling the mediating role of travel behavior.
Compared to \cite{luz2022understanding}, which treats achieved functionings as a single construct encompassing both travel behavior and activity participation, we decompose this stage into two distinct components: mobility (realized travel) and activity participation (Figure \ref{fig:concept}).
This decomposition allows us to model mobility as a mediating mechanism through which the capability set is selectively converted into actual leisure engagement, enabling identification of both direct and indirect pathways from accessibility to participation. \par

Transport planning approaches that equate accessibility solely with the provision of infrastructure or services often overlook these individual-level conversion processes, which are shaped by personal characteristics, environmental barriers, and cultural norms \citep{pereira2017distributive, van2022disentangling, luz2022understanding}.
This capability-oriented framework thus shifts attention to how varying levels of space–time accessibility are necessary to ensure equal opportunities, social inclusion, and the freedom to engage in third activities essential for well-being and development. \par

The flexibility and substitutability of leisure activities have direct implications for how we operationalize STA.
Because leisure destinations can be swapped, rescheduled, or foregone entirely, the relevant capability set is not a single location but a field of opportunities within an individual's time-space prism.
Someone with many reachable options can substitute one destination for another under time pressure; someone with a narrow set faces a binding constraint that may suppress participation altogether.
Moreover, because leisure occupies the residual time window after mandatory commitments, the STA value determines not only where individuals can go but whether they participate at all, making leisure activities relatively sensitive to variations in space-time resources. \par

In this study, we hypothesize that the factors shaping an individual's participation in leisure activities are the individual conversion function, the capability set, and the mobility, indicated by the thick green arrows in Figure \ref{fig:concept}.
The list of all opportunities accessible by a person (their \emph{capability set}) is operationalized from two building blocks: individuals' home and work/study locations, and their transport and amenity resources.

\section{Materials}\label{sec:materials}
This study draws on a combination of large-scale mobility traces, contextual geographic data, and socio-economic information to characterize individuals' opportunities and participation in activities in the Greater Paris region (Section \ref{sec:area}).
We integrate high-resolution GPS travel diaries from the 2023 Enquête Mobilité par GPS (EMG) (Section \ref{sec:data}) with detailed attributes on household and personal characteristics (Section \ref{sec:attr}), public transport and road network infrastructure (Section \ref{sec:network}), and point-of-interest (POI) distributions (Section \ref{sec:poi}).
Together, these materials provide a comprehensive foundation for quantifying space–time accessibility and examining the determinants of leisure and third-place activity participation across diverse population groups.

\subsection{Study area: The Paris region}\label{sec:area}
The Paris region, commonly known as the Île-de-France, is the most populous of France's eighteen regions, with an estimated 12.27 million residents as of January 2023 \citep{insee_pop_2020_regions}.
Centered on the capital Paris in the north-central part of the country, it covers 12,012 km$^2$, about 2\% of metropolitan France, yet accounts for nearly 20\% of the national population.
In the Paris region, population density reaches about 20,200 inhabitants/km$^2$ in the city of Paris, decreases to around 7,000 inhabitants/km$^2$ in the inner suburbs, and averages roughly 3,700 inhabitants/km$^2$ across the wider conurbation \citep{lagrandeconversation2023}. \par

\begin{figure*}[!ht]
\centering
\includegraphics[width=0.97\linewidth]{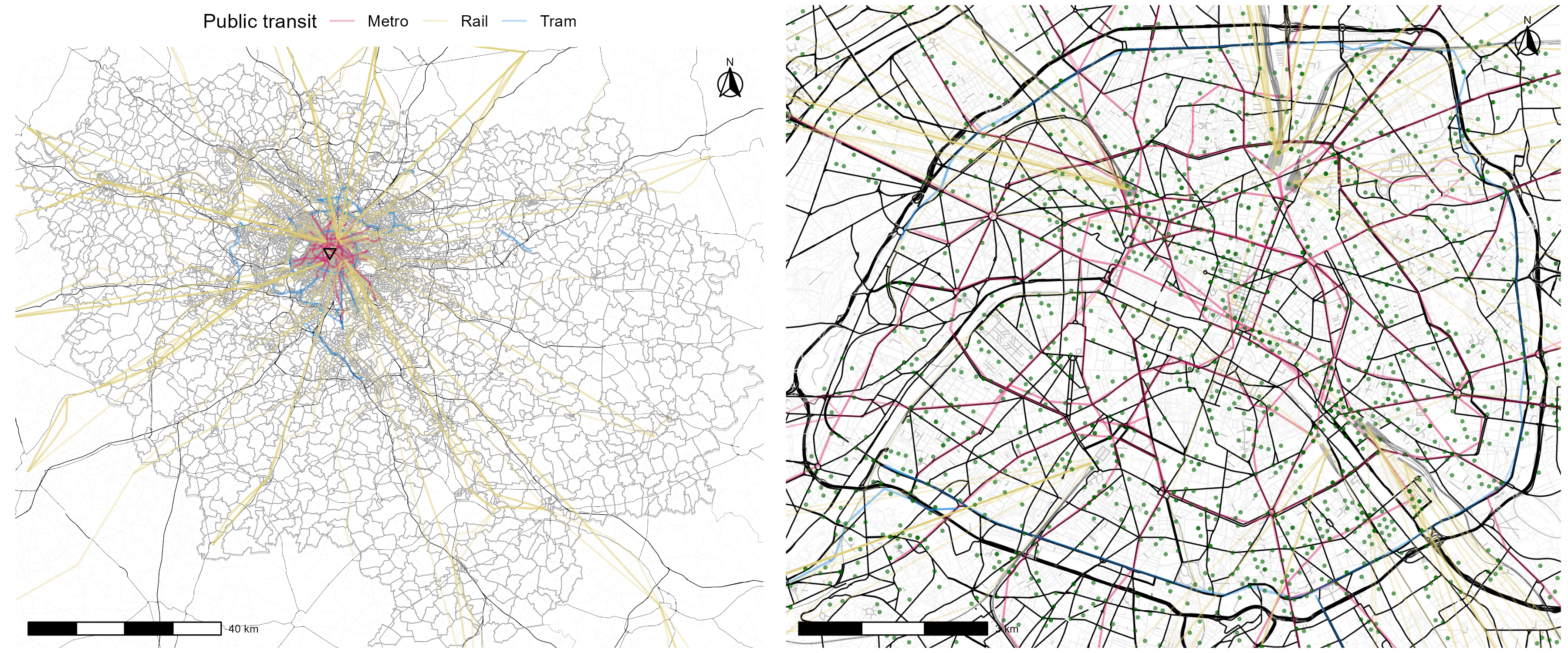}
\caption{\textbf{The Paris region and its public transit lines.} Grey lines represent IRIS boundaries, and black lines depict roads and motorways. The right panel provides a zoomed-in view of central Paris, where green points mark home and work locations.}
\label{fig:paris}
\end{figure*}

The Paris region (see Figure \ref{fig:paris}), with its monocentric urbanization pattern, has one of the world's densest and most multifaceted transport networks—spanning multiple transit modes and innovative services (metro, RER, tramways, buses, and on-demand services), coordinated by Île‑de‑France Mobilités (IDFM).
The region also offers robust ecosystem support via Mobility-as-a-Service (MaaS) platforms, integrating public transit with shared mobility, MaaS apps, and seamless ticketing \citep{idfmmob_mobility_as_a_service_2023}.
Residents frequently take multimodal trips and exhibit rich variation in their mobility patterns across individuals \citep{yin2023multimodal}.

\subsection{Trip records}\label{sec:data}
We use a dataset collected via the Enquête Mobilité par GPS (EMG 2023) initiative and made available through participating in the NetMob 2025 Data Challenge \citep{netmob25}.
The EMG 2023 survey, conducted between October 2022 and May 2023 covered 3,337 residents aged 16--80 in the Paris region, excluded non-residents, tourists, and the immobile, and used a multichannel recruitment strategy combining quotas and random draws to ensure representativeness \citep{netmob25}. \par

In this study, we analyze 14,169 individual-day records from 2,415 individuals with identified home and work locations, including information on date, day type (e.g., strike, holiday), origin, destination, transport mode, trip duration, and purpose.
Origins and destinations are represented by the centroids of the visited locations' corresponding H3 hexagons at resolution 10 \citep{uberh3}, each covering approximately 0.015 km$^2$. \par

Participants were observed for a median of 5 weekdays per individual (mean = 4.4, range 1–6).
The analysis was restricted to weekday observations (Monday–Friday), comprising 78.7\% of all recorded trips.
Although the raw data flagged atypical day types (public holidays, school holidays, bridge days, and strike days), we retained all weekday observations because the survey-provided day weights that adjust for representativeness across day types and because trips in such atypical days accounted for only 1.3\% of individual-weekday records. Individual-level variables were aggregated from daily records using weighted medians for individual metrics.

\subsection{Socio-economic attributes}\label{sec:attr}
The data from the EMG 2023 survey provides a rich set of sociodemographic, household, and mobility-related attributes.
It includes basic identifiers and residence information such as municipality codes, as well as demographic variables like sex, age, education level, and socio-professional category.
We also augment the attributes with the poverty rate at the IRIS (9-digit) zone level, used as a proxy for income in 2021.
This measure corresponds to the share of individuals living in households whose standard of living—after accounting for taxes, transfers, and social benefits—falls below 60\% of the national median disposable income \citep{INSEE_RevenusPauvreteIris2021}.
Household composition is captured through the number of persons and their age distribution, as well as the type of housing.
Mobility resources and constraints are represented by indicators of driving licence ownership and car availability, complemented by information on access to two-wheelers, bicycles, e-scooters, and other mobility devices.
The dataset also records subscriptions to public transport and other mobility services, alongside a statistical weighting coefficient for representativeness. \par

We focus on the main attributes: age, gender, poverty rate (zone level), education, household structure, main transport mode, active mode use, and public transport subscription.
Among these attributes, household structure captures gradations in domestic responsibility that shape time availability for leisure.
Single parents bear sole caregiving burden, representing the most constrained category.
Couples with children share significant obligations, typically involving escort trips and coordinated schedules.
By contrast, individuals living alone or in couples without children face minimal domestic care demands.
Those living with parents—often younger adults—may benefit from shared household tasks, partially offsetting their own time constraints.

\subsection{Public transit and road network data}\label{sec:network}
For public transit, we collected GTFS schedules from transport.data.gouv.fr, the French national open data platform for mobility, which provides official timetables and service information across transit operators (IDFM, accessed on 30 June 2025).
Road network data were obtained from OpenStreetMap, an open, collaboratively maintained geospatial database offering detailed and regularly updated representations of streets, pathways, and transport infrastructure (accessed on 1 July 2025).

\subsection{Point of interest (POI) data}\label{sec:poi}
POI data were collected via the Overture API, restricted to the geographical extent of the trip records, and filtered for a confidence level above 0.7 \citep{overturemaps}.
We then extracted POIs and their primary and secondary labels that are aligned with the activity purposes recorded in the trip data, \textit{Social \& Leisure} (67 k), consolidated through a combination of GPT-4o assisted classification (Essential needs, Health services, Education, Civic and utility, Social \& Leisure, and Other) and manual validation.

\section{Methodology}\label{sec:methods}
In this section, we outline the key components of the four methods used in our analysis.
We begin by describing the calculation of space–time accessibility (Section \ref{sec:sta}), followed by the approaches used to assess how well individuals' capability sets (reflected in their space–time accessibility) align with their observed spatial choices of leisure destinations (Section \ref{sec:selectivity}).
We then present the methods for quantifying mobility individually through total travel time and individual participation in third-place activities (Section \ref{sec:quantification}).
Finally, we introduce the structural equation model and its specification to disentangle the complex relationship between space-time accessibility and leisure activity participation (Section \ref{sec:sem}).
The source code of this study is available on \href{https://github.com/TheYuanLiao/netmob25}{GitHub}.

\subsection{Space-time accessibility}\label{sec:sta}
In this study, we operationalize individual-based capability set as \textit{space--time accessibility} (STA).
Specifically, we focus on the
\textit{cardinal accessibility} dimension, which accounts for the number of opportunities (leisure POIs, including parks, etc.) that individuals can feasibly reach within their available time budget \citep{saraiva2022accessibility}.
In other words, it counts the number of opportunities located within a person's Potential Path Area (PPA), which is the geographical area a person can reach given her space-time constraints \citep{saraiva2022accessibility}. \par

\begin{figure}[!ht]
\centering
\includegraphics[width=0.7\linewidth]{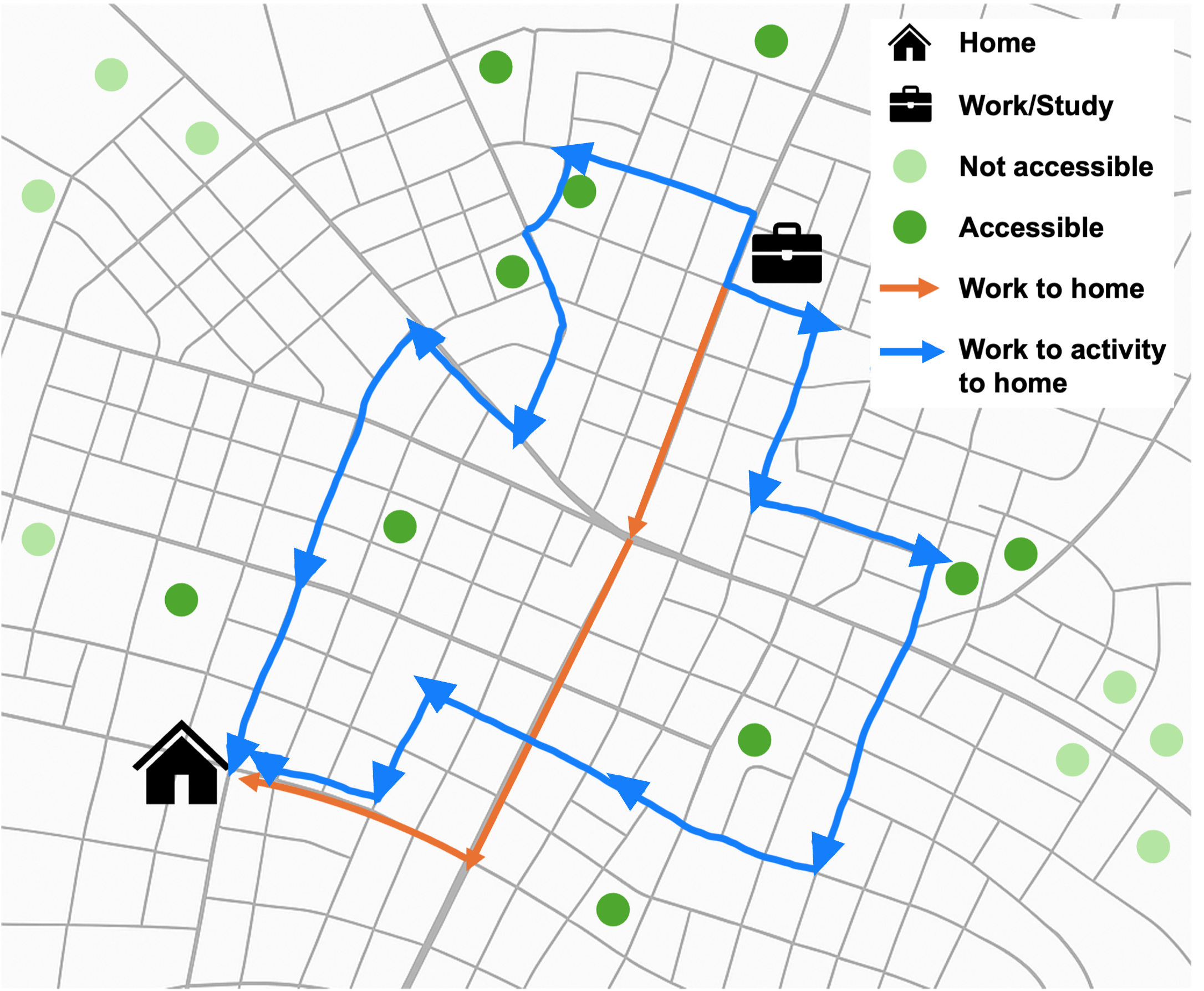}
\caption{\textbf{Space–Time Accessibility (STA)}. The light-green region represents the Potential Path Area (PPA), i.e., all locations reachable given the individual's time budget and travel constraints between home and work.
Opportunities within this area (dark green) are accessible, while those outside it (light green) are not.
The direct home–work path is shown in bold orange, and a feasible work–activity–home path is shown in blue.
This illustrates the STA of an individual, defined as the set of feasible opportunities that fall within the individual's PPA. Adapted from \cite{saraiva2022accessibility}.}
\label{fig:ak}
\end{figure}

The measure builds on the space-time feasibility concept \citep{kwan1999gender} and is modified based on the cardinal individual-based accessibility measure proposed in \citep{saraiva2022accessibility}.
Each individual is assumed to have a fixed total travel time budget $t_i^b$ and use their revealed main transport mode (car or public transport).
An activity destination $k$ is considered accessible to individual $i$ if the time required to visit $k$ does not exceed the individual's time budget.
In other words, the total time required to travel from home to work ($t_{hw,i}$), from work to $k$ ($t_{wk,i}$), and return from $k$ to home ($t_{kh,i}$) does not exceed the time budget, as if he/she does a one-stop trip chaining in between their trip from work to home, as more commonly found in the population \citep{mcguckin1999examining}.
In other words, the total time required to (i) travel from home to work ($t_{hw,i}$), (ii) travel from work to $k$ ($t_{wk,i}$), and (iii) return from $k$ to home ($t_{kh,i}$) does not exceed the time budget.

\begin{equation}
a^k_i =
\begin{cases}
1, & \text{if } t_{hw,i} + t_{wk,i} + t_{kh,i} \leq t_i^b \\
0, & \text{otherwise}
\end{cases}
\end{equation}%
where:
\begin{itemize}
    \item $t_{hw,i}$ is the travel time from home to work,
    \item $t_{wk,i}$ is the travel time from work to activity location $k$,
    \item $t_{kh,i}$ is the travel time from activity location $k$ to home,
    \item $t_i^b$ is the travel time budget of individual $i$.
\end{itemize}

With the overall space--time accessibility (STA$_i$, the capability set), the number of opportunities within STA$_i$ is $A_i$ for individual $i$, given by the sum of all feasible opportunities under certain category:

\begin{equation}
A_i = \sum_{k \in \mathcal{L}} a_i^k
\end{equation}
where $\mathcal{L}$ denotes the set of all candidate leisure POIs in the study area. \par

By construction, STA is mechanically determined by three inputs: the geolocations of home and work anchors, the travel speed of the main transport mode, and the time budget.
Variation across individuals thus arises not from the formula—which is identical for all—but from variation in these inputs, which are themselves socially structured: residential and workplace locations reflect housing markets and employment access; mode choice is shaped by income, household composition, and local infrastructure; and time budget is largely affected by the conjugation of activity responsibilities.
STA captures the spatial-temporal opportunity space that emerges from these socially determined conditions, consistent with the capability approach, in which conversion factors shape the resources that produce capabilities rather than entering the capability metric directly. \par

In this study, we set each individual's travel time budget $t_i^b$ to 90 minutes, based on the empirical median of 82 minutes observed in our dataset.
This threshold is consistent with the well-established travel time budget regularity i.e., that daily travel time budgets tend to cluster around 60–90 minutes across diverse urban settings, suggesting a stable behavioral constant in human mobility \citep{marchetti1994anthropological,schafer2000future}.
The fixed time budget is an intentional design choice rather than a limitation: adopting person-specific budgets derived from observed travel times would introduce circularity, as the explanatory variable would then be partly determined by the behavioral outcome it is meant to predict.
Commuting time ($t_{hw}$), travel time from home to work, is empirically estimated using the weighted median of observed commuting trips.
Discretionary activity $k$ is defined as a visit to leisure-related locations (e.g., bars). \par

For calculating $t_{wk}$ and $t_{kh}$, we fix the departure time at 17:00, which resembles the work-to-home trip chaining scenario with an added leisure activity stop on the way from work to home.
Travel times between home, work, and candidate leisure POIs were computed for both public transit and car, using GTFS schedules and road network data, implemented through the \texttt{r5r} package \cite{r5r}.
We also conduct and report sensitivity analysis with alternative budgets (i.e., 60, 75, 90, and 120 minutes) for both car and public transit and alternative departure times (i.e., 16:00, 17:00, and 18:00) for public transit. \par

Due to the presence of zero values and the right-skewed distribution of raw accessibility counts, we apply the inverse hyperbolic sine (IHS) transformation:
\begin{equation}
  A_i^* = \sinh^{-1}(A_i) = \ln\left(A_i + \sqrt{A_i^2 + 1}\right)
\end{equation}
where $A_i$ is the raw count of accessible leisure POIs for individual $i$, and $A_i^*$ is the transformed value used in the downstream analysis.
The IHS transformation behaves similarly to the natural logarithm for large values, but is defined at zero ($\sinh^{-1}(0) = 0$), which preserves the zero-accessibility cases while reducing the influence of extreme values. \par

Because the STA framework requires both a home and a work anchor to define the space–time prism, our analysis is restricted to those for whom both locations could be identified, effectively limiting the sample to the working population and excluding non-workers, retirees, and unemployed individuals whose accessibility constraints may differ substantially.

\subsection{Capability set vs. spatial choices} \label{sec:selectivity}
To gain a finer-grained understanding of how accessibility shapes behavior, we assess whether individuals tend to select activity locations that align with what their modeled opportunity landscape (STA) predicts as desirable, or whether actual behavior departs from this logic. \par

\begin{figure}[!ht]
\centering
\includegraphics[width=0.7\linewidth]{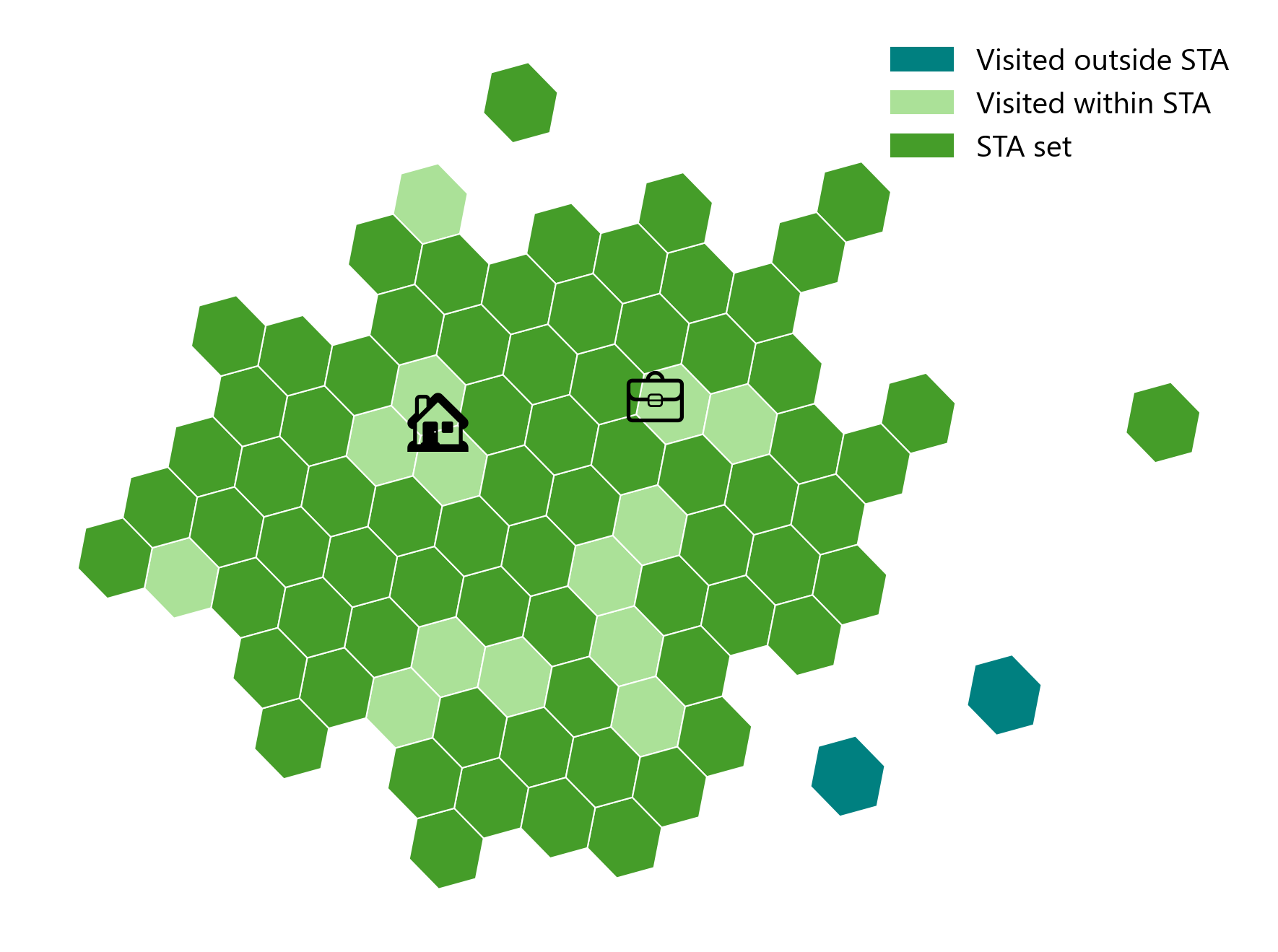}
\caption{\textbf{Space–Time Accessibility (STA) set vs. visited leisure locations}.
One example individual.
Home and work icons show where the person lives and works. Hexagons are at resolution 8, $\sim$0.74 km$^2$.}
\label{fig:ak-example}
\end{figure}

This behavioral analysis focuses on the alignment between the ranked set of feasible locations STA$_i$ (adding up to $A_i$) within individuals' potential path area and the locations they actually visited.
By comparing observed choices against random draws from STA$_i$, we test for selectivity---that is, whether individuals systematically prefer most accessible locations (see Figure \ref{fig:ak-example}). \par

We evaluate whether individuals tend to visit the activities that are the most accessible (e.g., those that take a shorter time to reach) using a ranked comparison:
For each individual $i$, we operationalize their STA$_i$ as a ranked list of feasible locations (H3 hexagons at resolution 8, $\sim$0.74 km$^2$).
We focus on H3 hexagons rather than raw GPS coordinates for two reasons.
First, not all locations labeled as leisure in the GPS data correspond to entries in the leisure POI dataset extracted from Overture Maps.
Second, the GPS coordinates have been preprocessed by the data provider to H3 level-10 centroids as a privacy measure, making precise point-level POI matching infeasible.
Consequently, POI assignment must rely on an appropriate spatial unit rather than exact-coordinate matching.
These spatial units (H3 hexagons at resolution 8) are further sorted from 1 (best = most accessible, lowest travel time increase) to $N_i$ (worst = least accessible, highest travel time increase). \par

We then compare the set of visited locations $V_i$ to this ranking.
First, we calculate the share of locations in $V_i$ that fall outside STA$_i$.
Second, we define the test statistic as the average rank of the visited locations falling within STA$_i$:
\begin{align}
T_{\text{mean},i} = \frac{1}{K_i} \sum_{\ell \in V_i} \text{rank}(\ell)
\end{align}
where $K_i = |V_i|$. \par

To generate a null distribution, we perform $B=1000$ random draws of $K_i$ locations from STA$_i$, computing $T_{\text{mean},i}^{(b)}$ for each.
We then derive both the empirical $p$-value:
\begin{align}
p_i = \frac{1 + \#\left\{ b : T_{\text{mean},i}^{(b)} \leq T_{\text{mean},i} \right\}}{B + 1}
\end{align}%
and the standardized effect size:
\begin{align}
d_i = \frac{T_{\text{mean},i} - \mu_i^{\text{null}}}{\sigma_i^{\text{null}}}
\end{align}%
where $\mu_i^{\text{null}}$ and $\sigma_i^{\text{null}}$ are the mean and standard deviation of the null distribution.
Negative values of $d_i$ indicate better-than-random performance---i.e., that the individual tends to visit higher-ranked locations than would be expected under random choice.

\subsection{Weekday mobility and activity participation}\label{sec:quantification}

We consider total travel time for \emph{mobility} quantification.
For \emph{activity participation}, we consider two aspects: 1) the diversity of locations where activities occur (diversity), and 2) the mean leisure activity duration (intensity).
Because the travel diary lacks detailed labeling of third-place activities, we approximate third-place participation using all leisure activities ('LEISURE'). \par

We compute the diversity measure based on the distribution of leisure activity locations.
For each individual $i$, we count the number of visits to each distinct leisure location (H3 hexagons at resolution 10) during their data collection period.
Let $n_i$ denote the total number of leisure visits, and let $K^d_i$ denote the number of distinct leisure locations visited.
Defining $p_{k,i} = n_{k,i} / n_i$ as the relative frequency of visits to leisure location $k$ ($k=1,\dots,K^d_i$),
we compute the \emph{Hill number of order $q=1$}, which is given by:

\begin{equation}
H_{1i} = \exp\left( - \sum_{k=1}^{K^d_i} p_{k,i} \, \ln(p_{k,i}) \right).
\end{equation}%
which reflects the effective number of equally frequent locations visited, and accounts for both richness and evenness in the individual's leisure location distribution. \par

Location diversity and mean leisure duration are assessed to be valid indicators of the common latent construct, leisure activity participation.
The two measures share moderate positive correlation (corr. efficient = 0.31, weighted), and Bartlett's test confirmed factor analysis is appropriate ($\chi^2$ = 208.7, p < 0.001).
A single-factor extraction yielded equal loadings of 0.54 for both indicators, above the conventional 0.40 threshold.
The modest shared variance (28.6\%) reflects their complementary nature: diversity captures the breadth of leisure engagement across locations, while duration captures depth of time invested.
These test outcomes justify their treatment as parallel indicators rather than causally ordered outcomes. \par

In operationalizing the quantification of mobility and activity participation, we focus exclusively on leisure activities during weekdays.
Since our space-time accessibility measure is built on the home–work–activity–home trip chain, restricting the analysis to regular weekdays ensures consistency between the assumptions underlying the STA metric and the observed travel behavior used to measure mobility and leisure participation.
Individual-level indicators (total travel time, leisure location diversity, mean duration of leisure activities) were aggregated across all valid weekday observations for each person.

\subsection{Structural equation modeling specification}\label{sec:sem}

Based on the data summarized in Table \ref{tab:desc_stats}, we estimate a structural equation model (SEM) linking (i) individual attributes, (ii) transport mode, (iii) space-time accessibility value $A_i^*$, (iv) total travel time, and (v) leisure activity participation (latent variable).
Let $i$ index individuals. The observed variables are:

\begin{itemize}
  \item[i.] $A_i^*$: IHS--transformed space--time accessibility to leisure opportunities;
  \item[ii.] $\mathbf{Z}_i$: a vector of exogenous sociodemographic and lifestyle attributes (dummy--coded for household type, gender, education, etc.), retained after preprocessing to remove colinearity and low variance;
  \item[iii.] $\mathbf{M}_i$: transport variables, including the main transport mode dummy and the public transport subscription dummy;
  \item[iv.] $B_i$: travel behavior indicator, namely total travel time;
  \item[v.] $\eta_i$: leisure activity participation, measured by the Hill number of order $q=1$ ($H_{1i}$), quantifying the diversity of leisure locations visited, and mean leisure activity duration in minute ($T_i$), quantifying the activity intensity.
\end{itemize}

\begin{table*}[htbp]
\centering
\caption{\bf Descriptive statistics of the 2,415 commuting individuals. Values represent weighted means (continuous variables) and weighted percentages (categorical variables). $A_i$ is transformed using the inverse hyperbolic sine (IHS), while zeros are retained.}
\begin{tabular}{clll}
\hline
\multicolumn{1}{l}{Group} & Variable & Levels & Mean (SD) or \% \\\hline

\multirow{22}{*}{Individual attributes}
& Age & - & 43.3 (11.8) \\\cline{2-4}
& Poverty rate (IRIS zone level) & - & 15.8 (8.3) \\\cline{2-4}

& \multirow{7}{*}{Education}
& No diploma & 1.7 \\
& & Vocational & 11.7 \\
& & Lower secondary & 3.0 \\
& & Upper secondary & 31.4 \\
& & 3--4-year higher education & 15.7 \\
& & 5-year-and-above higher education & 25.6 \\
& & Missing & 10.9 \\\cline{2-4}

& \multirow{2}{*}{Gender}
& Man & 47.6 \\
& & Woman & 52.4 \\\cline{2-4}

& \multirow{8}{*}{Household type}
& Living alone & 12.2 \\
& & In a couple w/o children & 20.3 \\
& & Single parent & 11.6 \\
& & Living with parent(s) & 6.2 \\
& & Not related to other household members & 1.1 \\
& & In a shared apartment & 0.3 \\
& & In a couple w/ child(ren) & 47.2 \\
& & Another family member in the household & 1.1 \\\cline{2-4}

& \multirow{2}{*}{Use of active mode}
& No & 62.1 \\
& & Yes & 37.9 \\\hline

\multirow{3}{*}{Space-time accessibility}
& \multirow{2}{*}{STA value (\textgreater 0)}
& Non-zero & 39.7 \\
& & Zero & 60.3 \\\cline{2-4}
& STA value (IHS) & - & 3.2 (4.1) \\\hline

\multirow{4}{*}{Transport mode}
& \multirow{2}{*}{Main transport mode}
& Car & 38.7 \\
& & Public transit & 61.3 \\\cline{2-4}
& \multirow{2}{*}{Public transit subscription}
& No & 35.7 \\
& & Yes & 64.3 \\\hline

Mobility
& Total travel time (min) & - & 96.2 (44.1) \\\hline

\multirow{2}{*}{Activity participation}
& Diversity & - & 1.1 (1.4) \\
& Duration (min) & - & 62.5 (99.4) \\\hline
\end{tabular}
\label{tab:desc_stats}
\end{table*}

To avoid imposing causal ordering between two parallel outcomes, we tested whether location diversity and leisure duration could serve as indicators of a latent ``leisure participation'' construct.
The two variables are significantly correlated ($r = 0.31$, $p < 0.001$), and Bartlett's test confirmed shared variance ($\chi^2 = 208.7$, $p < 0.001$).
Exploratory factor analysis yielded adequate loadings for both indicators ($\lambda = 0.54$), exceeding the 0.4 threshold for meaningful loadings \citep{hair2010multivariate}.
We therefore adopt a latent variable specification in the SEM. \par

Before estimating the structural equation model, it is important to specify a directed acyclic graph (DAG) that encodes the theoretical and causal assumptions about the relationships among the variables.
The DAG serves as a transparent foundation for identifying potential confounding paths, making the assumptions of the model explicit, and ensuring that the SEM structure is both theoretically grounded and empirically justified \citep{huntington2021effect}.
We iteratively evaluated which relationships could be pruned by testing the key conditional independence implications suggested by our theoretical framework.
We assessed whether certain connections between variables---such as those linking individual attributes, transport mode, capability set, mobility, and activity participation---could be removed without contradicting the observed data.
This process allowed us to identify which dependencies were empirically warranted and which were not, ensuring that the resulting DAG retained only the pathways supported by theory and evidence. \par

The final directed acyclic graph (DAG) reflects the theoretically informed and empirically tested structure of the relationships among individual attributes, transport mode, capability set, mobility, and activity participation (Figure~\ref{fig:dag_final}).
Individual characteristics are modeled as exogenous factors that shape accessibility (capability set), as well as exerting direct influences on trip-making and activity participation.
Transport mode variables are also treated as exogenous.
Modeling mode choice endogenously led to non-convergence due to identification issues arising from the relationship between binary mode indicators and continuous outcomes.
Moreover, mode choice depends on complex factors that are better addressed through dedicated discrete choice models \citep{ben1985discrete}.
Our focus is on how transport mode conditions accessibility and leisure participation, not on the determinants of mode choice itself.
Therefore, covariances between sociodemographic and transport mode variables are freely estimated, preserving their correlational structure without imposing a causal ordering.
The capability set, representing the feasible space–time leisure opportunities available to individuals, is affected by both personal attributes and transport mode, and in turn drives patterns of trip-making and participation.
Trip-making operates as an intermediate behavioral mechanism linking accessibility to leisure activity participation.
Finally, leisure activity participation is conceptualized as the outcome of multiple interacting pathways: directly from individuals, from transport behavior, and indirectly via both the capability set and trip-making.
This structure balances theoretical expectations from time-geography and accessibility theory with empirical evidence from conditional independence testing, ensuring that only statistically supported pathways are retained. \par

\tikzset{
  node/.style={draw, rounded corners, align=center, minimum width=2.2cm, minimum height=1.0cm},
  exo/.style={node, fill=Exo},
  med/.style={node, fill=Med},
  beh/.style={node, fill=Beh},
  outc/.style={node, fill=Outc},
  arrow/.style={-{Latex}, thick},
  causal/.style={arrow, draw=black},
  bias/.style={arrow, draw=Bias, thick},
}

\begin{figure}[!ht]
\centering
\begin{tikzpicture}[node distance=1.5cm and 2cm]

\node[exo] (indiv) {Individual\\attributes\\($\mathbf{Z}_i$)};
\node[exo, right=of indiv] (mode) {Transport\\mode\\($\mathbf{M}_i$)};
\node[med, below=of indiv] (capset) {STA value\\($A_i^*$)};
\node[beh, right=of capset] (trips) {Mobility\\($B_i$)};
\node[outc, below=of trips] (part) {Activity\\participation\\($\eta_i$)};

\draw[bias] (indiv) -- (capset);
\draw[bias] (indiv) -- (trips);
\draw[bias] (indiv) -- (part);

\draw[bias] (mode) -- (capset);
\draw[bias] (mode) -- (trips);
\draw[bias] (mode) to[bend left=40] (part);

\draw[causal] (capset) -- (trips);
\draw[causal] (capset) -- (part);

\draw[bias] (trips) -- (part);

\end{tikzpicture}
\caption{\textbf{Pathway structure}.
This directed acyclic graph shows the hypothesized and empirically validated pathways.
Exposure=STA value, Outcome=Activity participation, Black arrows=Causal paths, Brown arrows=Biasing paths.}
\label{fig:dag_final}
\end{figure}

The SEM specifies the relationships in Figure~\ref{fig:dag_final}.
In the measurement model, the latent leisure participation construct $\eta_i$ is measured by two observed indicators:
\begin{align}
H_{1i} &= \lambda_1 \eta_i + \epsilon_{1i}, \\
T_{i} &= \lambda_2 \eta_i + \epsilon_{2i},
\end{align}
where $H_{1i}$ is location diversity, $T_{i}$ is mean leisure duration, $\lambda_1 = 1$ (fixed for identification), and $\lambda_2$ is freely estimated. \par

The structural model is shown below:
\begin{align}
  A_i^* &\sim \boldsymbol{\alpha}_Z^\top \mathbf{Z}_i + \boldsymbol{\alpha}_M^\top \mathbf{M}_i, \\
  B_i &\sim a \, A_i^* + \boldsymbol{\gamma}_Z^\top \mathbf{Z}_i + \boldsymbol{\gamma}_M^\top
  \mathbf{M}_i, \\
  \eta_i &\sim c \, A_i^* + b \, B_i + \boldsymbol{\delta}_Z^\top \mathbf{Z}_i +
  \boldsymbol{\delta}_M^\top \mathbf{M}_i,
\end{align}
where $A_i^*$ is space-time accessibility, $B_i$ is total travel time, $\mathbf{Z}_i$ is a vector of sociodemographic covariates (gender, education, poverty rate, household type), and $\mathbf{M}_i$ is a vector of transport mode variables (public transit mode, PT subscription, active mode). \par

The total effect of space-time accessibility on each observed indicator is:
\begin{align}
  \text{Total effect on } H_{1i} &= (c + a \cdot b) \times \lambda_1, \\
  \text{Total effect on } T_i &= (c + a \cdot b) \times \lambda_2,
\end{align}
where $c$ is the direct effect, and $a \cdot b$ is the indirect effect mediated through travel time. \par

The analysis was performed in R (\texttt{lavaan} 0.6-19) using the diagonally weighted least squares (DWLS) estimator, which is appropriate for models that include categorical or non-normally distributed variables.
Continuous variables were standardized to stabilize variances.
Weights were normalized to have a mean of one and applied during estimation to account for individual-level representativeness.
Multicollinearity was assessed using variance inflation factors (VIF), applying thresholds of 5 (moderate) and 10 (severe).
All retained predictors had VIF < 2.2. Education was recoded from 7 levels to a binary (high vs. low/medium) due to high correlation among the category dummies (r = -0.89).
Household type effects were included only in equations where preliminary estimation indicated statistical significance (p < 0.05); non-significant household effects were constrained to zero to improve model parsimony and identification.
Rare household categories (< 2\% of the sample) were excluded.
Gender, poverty rate, transport mode, public transit subscription, and active mode use were retained as predictors across all equations.
We report standardized coefficients, $R^2$ values for endogenous variables, and standard SEM fit indices (CFI, TLI, RMSEA, SRMR).
Indirect, direct, and total associations are computed through model-implied parameter combinations. \par

\section{Results and discussion}\label{sec:results}
In this section, we first visualize space–time accessibility and leisure activity participations (Section \ref{sec:res_desc}).
We then report the results assessing the alignment between the modeled capability set and individuals' actual spatial choices of leisure destinations (Section \ref{sec:res_selectivity}).
Finally, we present the modeling outcomes (Section \ref{sec:res_sem}), detailing the quantified relationships and how different components link to space–time accessibility and shape leisure location diversity.

\subsection{Space-time accessibility and leisure activity participation}\label{sec:res_desc}
\begin{figure*}[!ht]
\centering
\includegraphics[width=0.97\linewidth]{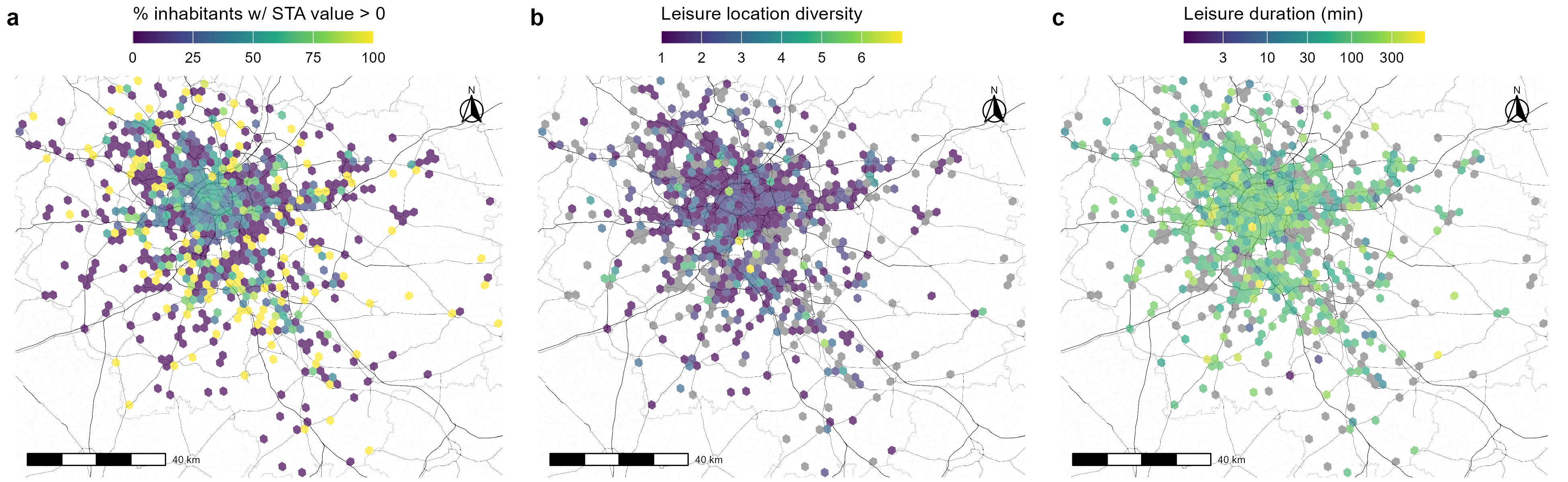}
\caption{\textbf{Space–time accessibility value and mobility metrics on the map.}
\textbf{a}, Share of inhabitants with $A_i^*$ > 0.
\textbf{b}, Leisure location diversity.
\textbf{c}, Leisure activity duration (min), on log-transformed colour scale.
For \textbf{b}-\textbf{c}, the map shows the weighted median of each zone's inhabitants in the data.}\label{fig:ak_tc}
\end{figure*}

Figure~\ref{fig:ak_tc} presents a spatial overview of space–time accessibility value $A_i^*$ and leisure activity participation measures across the study region, highlighting the geographic variations.
Figure~\ref{fig:ak_tc}a shows the proportion of inhabitants with STA~$>0$, indicating the share of individuals who have feasible opportunities beyond their fixed home and work locations.
Higher values are concentrated in the northern central urban areas and scattered across select suburban zones, suggesting greater access to leisure activity opportunities in both dense and well-connected peripheral locations, mainly by car.
Activity participation lacks clear spatial patterns (Figures~\ref{fig:ak_tc}b–c).
This suggests that leisure activity participation is more strongly driven by individual-level differences, underscoring the need for appropriate modeling.
Sensitivity analyses confirm that STA patterns are robust to variation in both the time budget (60–120 minutes) and the departure hour (16:00–18:00); details are reported in Appendix \ref{seca:sta} (Figure \ref{fig:sensitivity}). \par

\begin{figure}[!ht]
\centering
\includegraphics[width=0.7\linewidth]{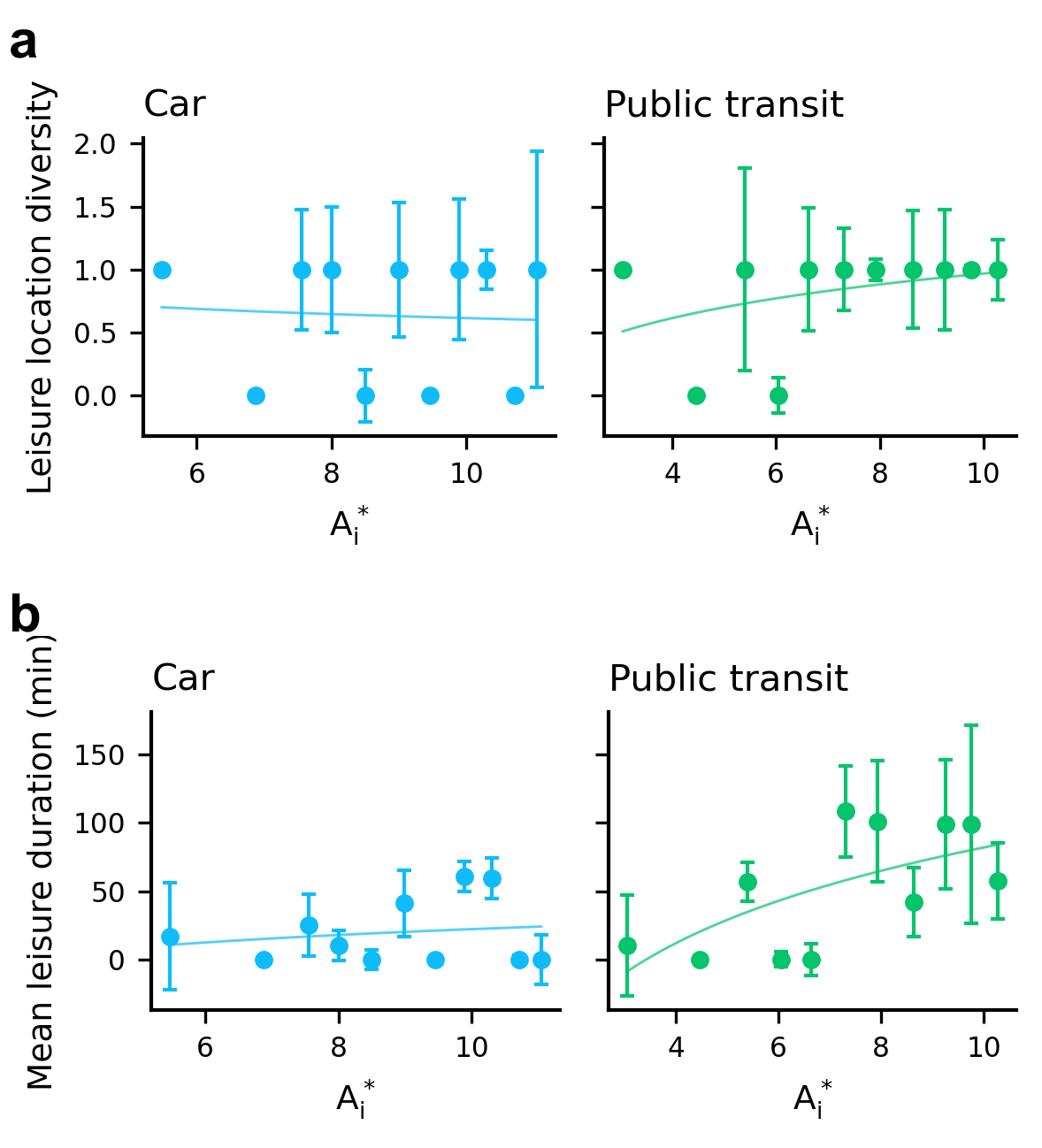}
\caption{\textbf{Space–time accessibility value $A_i^*$ vs. leisure activity participation by mode.}
\textbf{a}, Leisure location diversity.
\textbf{b}, Leisure activity duration (min).
Error bars indicate bootstrap median estimation errors.}
\label{fig:res1}
\end{figure}

We further visualize how STA relates to leisure activity participation, stratified by main transport mode (car vs. public transit) in Figure~\ref{fig:res1}.
Leisure location diversity and duration are in general positively associated with STA, especially for public transit users.
These visual patterns support our assumptions regarding the interdependence of individual attributes, transport mode choice, and space–time accessibility value (Figure \ref{sec:concept}).
SEM outcomes further disentangle their relationships.

\subsection{Selectivity in spatial choice of weekday leisure activities}\label{sec:res_selectivity}
Before examining the structural relationships between STA and leisure participation, we first assess whether the modeled capability sets align with individuals' actual spatial choices, which serves as a necessary validation step to ensure that the STA metric meaningfully captures the opportunity structures it is designed to represent. \par

\begin{figure}[!ht]
\centering
\includegraphics[width=0.7\linewidth]{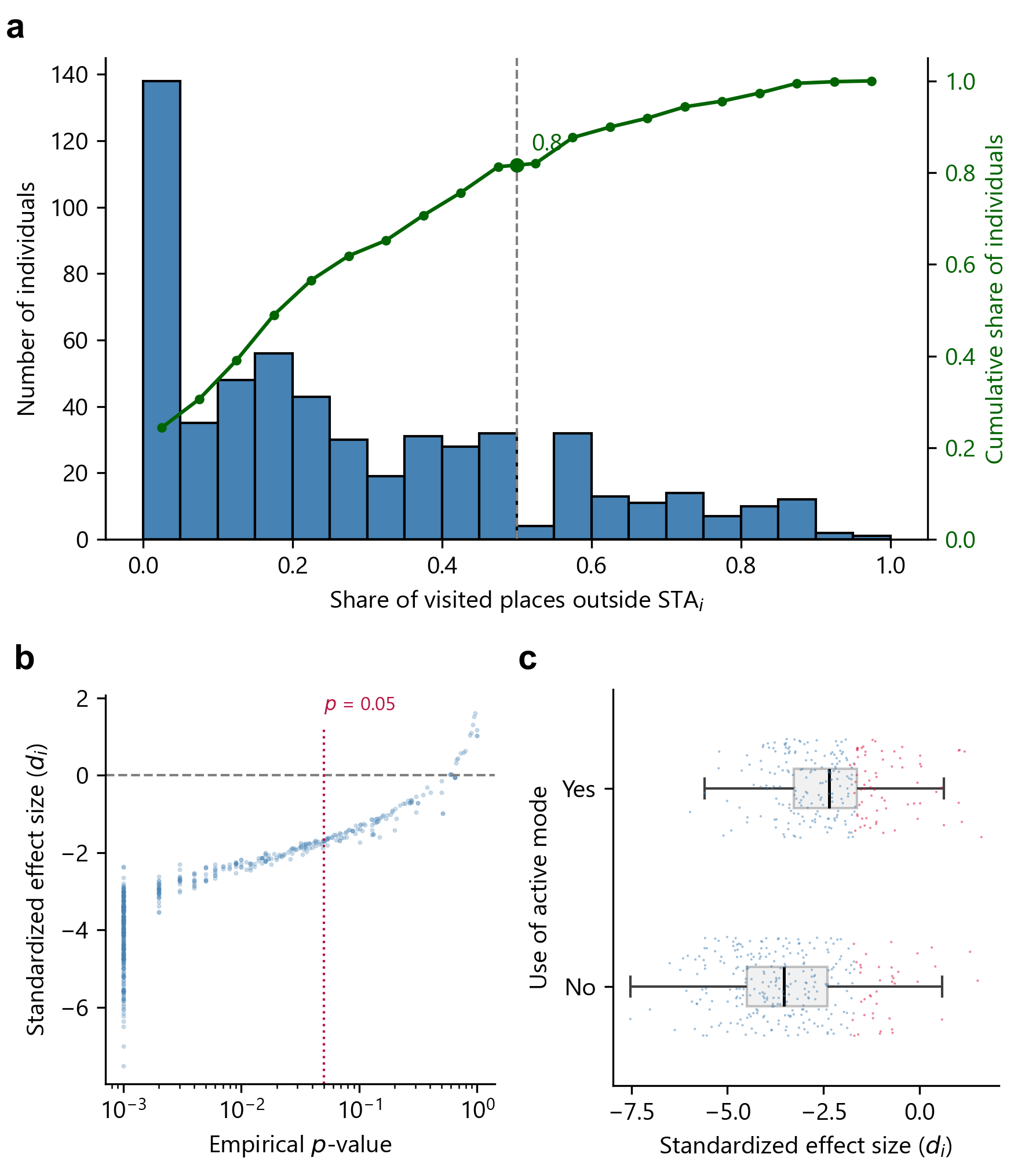}
\caption{\textbf{Selectivity in spatial choice behavior across individuals.}
\textbf{a}, Distribution of individuals by the share of visited locations outside their modeled feasible set (STA$_i$), with bar color indicating the proportion of car users in each bin. Bins include lower boundaries.
\textbf{b}, Standardized effect size $d_i$ plotted against the empirical $p$-value from the mean rank test, on a log scale. The vertical line marks the $p = 0.05$ threshold. Each point is an individual.
\textbf{c}, Boxplot of $d_i$ stratified by use of active transport mode, with jittered points colored by statistical significance (blue: $p < 0.05$).}
\label{fig:selectivity}
\end{figure}

Based on 566 individuals where their $A_i>0$ and have leisure activities registered during weekdays, we observe a strong alignment between STA$_i$ and actual spatial choices of leisure activities (Figure \ref{fig:selectivity}).
The distribution of visited locations outside STA$_i$ is skewed toward lower values (Figure \ref{fig:selectivity}a): the top 20\% of individuals make more than half of their visits outside their capability set.
Moreover, only 0.2\% of individuals have visited zero places within their capability set.
Space-time accessibility in this context primarily captures trip-chaining opportunities between home and work, framing leisure visits as secondary or incidental. \par

Most individuals exhibit negative $d_i$ values (Figure \ref{fig:selectivity}b), indicating a tendency to choose closer locations than expected by chance, as 80\% of individuals have significant deviation from being random ($p<0.05$).
This supports the feasibility of using STA as a realistic approximation of individuals' capability sets.
A cluster of individuals exhibits strongly negative $d_i$ values with highly significant $p$-values, while others have values near zero or even positive, suggesting behavior closer to random or worse than random.
These results reveal considerable heterogeneity in the degree of behavioral selectivity across the population. \par

Among all individual attributes considered, only the use of active transport mode shows a statistically significant association with the effect size ($d_i$), as illustrated in Figure~\ref{fig:selectivity}c.
Active users exhibit a higher median $d_i$ and a tighter distribution skewed toward weaker selectivity (i.e., more positive values).
Individuals who use active modes exhibit significantly higher $d_i$ than non-active travelers, with an estimated increase of 1.04 units ($p < 0.001$), suggested by the linear regression test $d_i\sim C(\text{Using Active Mode})$.
These results suggest that individuals using active mode tend to align more closely with their STA-based opportunity rankings, whereas car users may be less constrained or freer in their spatial choices. \par

Taken together, individuals exhibit varying degrees of alignment with modeled opportunity structures.
Active transport users show stronger within-STA selectivity.
These findings underscore the value of decomposing the behavioral mechanisms underlying accessibility: individuals do not uniformly act upon modeled opportunities, and their selectivity depends critically on their lifestyle and spatial constraints.

\subsection{Modeling outcomes}\label{sec:res_sem}
Having established that individuals' leisure choices align with their modeled capability sets, we now examine whether the breadth of these sets, i.e., space–time accessibility, predicts the diversity of leisure locations visited. \par

We apply structural equation modeling techniques with the diagonally weighted least squares (DWLS) estimator on a total of 2,415 weighted observations (individuals) to model leisure activity participation during weekdays.
The model demonstrated excellent overall fit, with a Comparative Fit Index (CFI) of 0.987 and Tucker--Lewis Index (TLI) of 0.954.
The Root Mean Square Error of Approximation (RMSEA) was 0.029, with a 90\% confidence interval of [0.018, 0.041], and the Standardized Root Mean Square Residual (SRMR) was 0.014, all supporting model adequacy \citep{hu1999cutoff}.
We assess the robustness of the SEM results to the time budget parameter by estimating the model under restrictive (60 min) and generous (120 min) specifications, fixing departure time at 17:00 given its limited impact (Figure \ref{fig:sensitivity}).
Model fit remains good across all specifications, and the pattern of effects are consistent (see Appendix \ref{seca:sem}).
Therefore, we report the detailed results from the 90-minute specification below.\par

\begin{figure*}[!ht]
\centering
\includegraphics[width=0.97\linewidth]{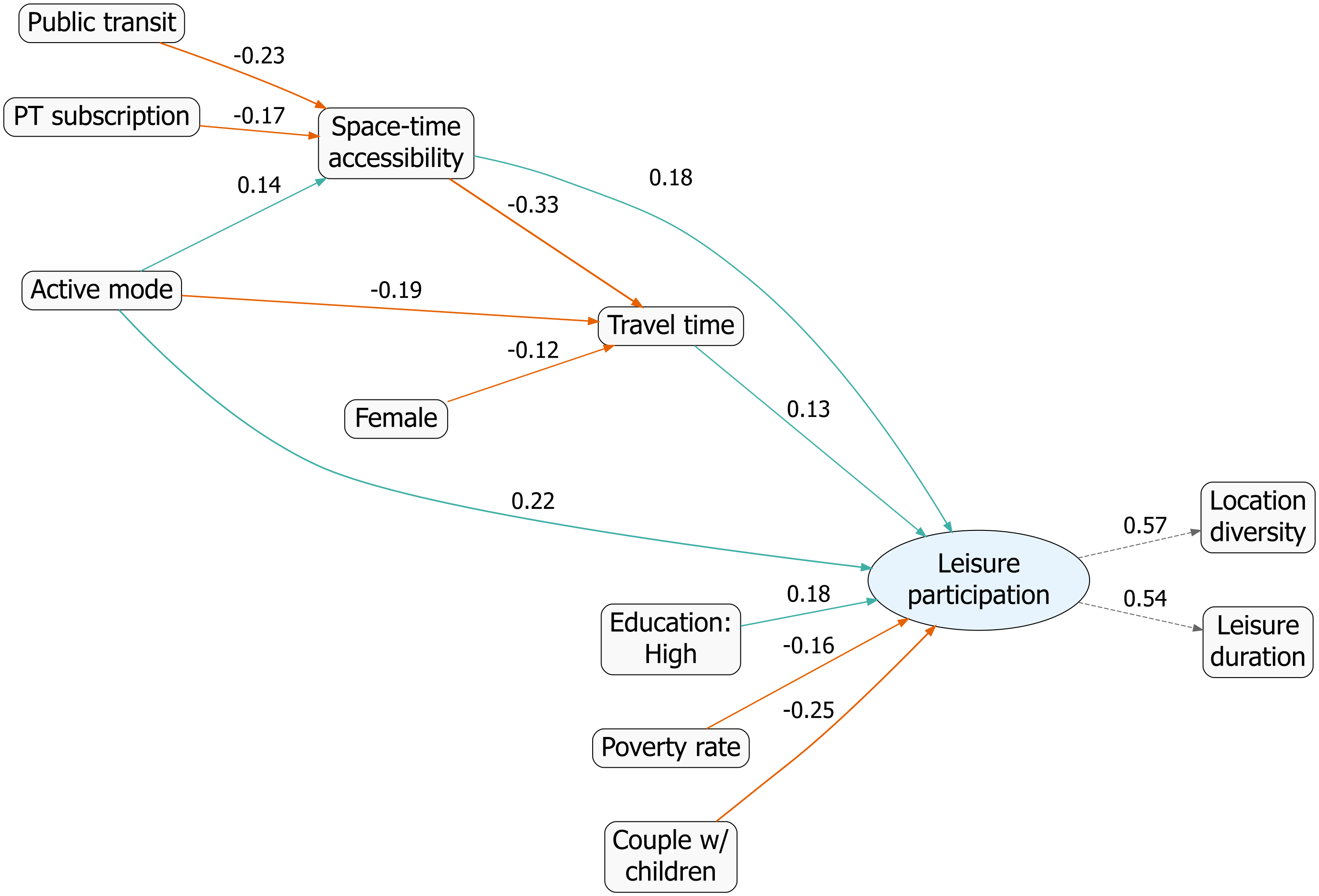}
\caption{\textbf{Structural equation model of factors shaping leisure activity participation.}
Standardized path coefficients ($p<0.05$) are shown along the arrows, with positive effects in teal and negative effects in orange.}
\label{fig:sem_plot}
\end{figure*}

Significant direct and indirect pathways with standardized coefficients exceeding an absolute value of 0.1 (Figure~\ref{fig:sem_plot}) highlight the role of household structure, active mode use, and public transport subscription in shaping space–time accessibility, total travel time, and leisure activity outcomes. \par

\subsubsection{Transport mode predicts space-time accessibility}
Space-time accessibility is shaped by transport characteristics.
Active mode users---those regularly walking or cycling---exhibit significantly higher STA value ($\beta = 0.14$, $p < .001$).
Given how STA is evaluated, this could be ascribed to two main mechanisms:
(1) these individuals mostly live in the city of Paris, with good \textbf{proximity} to leisure opportunities and good support for walking and cycling; and (2) they have relatively \textbf{short commuting times} (34 min) compared to non-active mode user (43 min), making the assumed 90-minute daily travel time budget have a greater share available for potential leisure activities. \par

We find that both public transit use ($\beta = -0.23$, $p < .001$) and public transport subscription ($\beta = -0.17$, $p < .001$) are negatively associated with STA.
This suggests that, given individuals' fixed daily anchors (home and work/study locations) and their chosen mode of travel, the car provides more effective access to a wider range of leisure opportunities compared to public transport. \par

All paths connecting individual attributes to STA are found to be insignificant, with their effects primarily on travel time and leisure participation.

\subsubsection{Space-time accessibility predicts travel time together with active mode usage and gender}
Space-time accessibility strongly predicts total travel time ($\beta = -0.33$, $p < .001$).
Good STA implies a greater number of accessible leisure opportunities within the time–space prism, which in turn leads to shorter travel times, indicating that part of the activity demand can be met in proximity.
The use of active modes is also associated with shorter total travel time ($\beta = -0.19$, $p < .001$), suggesting that those who rely on active transport do so because their activity needs can be met closer to their home and work anchors.
Females tend to have shorter travel times compared to males ($\beta = -0.12$, $p < .001$).
The total travel time, in turn, serves as a mediator between STA and leisure activity participation.

\subsubsection{Leisure activity participation predicted by space-time accessibility, travel time, and individual attributes}
Higher accessibility is positively associated with greater leisure participation ($\beta = 0.18$, $p < .001$).
This finding underscores a conceptual alignment between space-time accessibility to leisure opportunities and actual participation in diverse leisure locations (0.57) and longer duration (0.54).
In other words, individuals with higher accessibility to leisure opportunities are more likely to translate potential into visits to a broader range of leisure locations and longer activities. \par

Mobility also contributes here: longer travel times increase leisure location diversity ($\beta = 0.13$, $p = <.001$).
From the perspective of utility-maximization, longer travel time can still result in higher overall utility if it enables meaningful leisure engagement \citep{de2016travel}.
This effect is around the same magnitude as that of STA directly, underscoring the role of total travel time as a key mediator shaping actual leisure activity participation. \par

Active mode use is positively associated with leisure activity location diversity ($\beta = 0.22$, $p = <.001$).
Leisure travel is often motivated by social needs, the pursuit of variety, and personal recharge \citep{stauffacher2005diversity}.
Active transport may facilitate these motivations by providing greater flexibility in route selection and opening up opportunities to access varied leisure spaces such as parks, cafés, or community centers.
It's plausible that walking or cycling itself becomes part of a user's leisure experience---adding intentionality and value to the travel act. \par

Individual attributes also shape leisure participation directly.
Higher education is positively associated with leisure location diversity ($\beta = 0.18$, $p < .001$), consistent with evidence that education expands both awareness of and preference for varied leisure opportunities.
Conversely, higher local poverty rates ($\beta = -0.16$, $p < .001$) and living in couple-with-children households ($\beta = -0.25$, $p < .001$) are negatively associated with leisure participation, reflecting how economic deprivation and caregiving responsibilities constrain discretionary activity engagement beyond the indirect effects already captured through transport mode and accessibility pathways.

\subsubsection{Effect decompositions of space-time accessibility}
Decomposition analyses reveal that the space-time accessibility value had both direct and indirect effects on leisure participation (Table \ref{tab:effects}).
The direct effect of accessibility on leisure location diversity is positive and significant ($\beta = 0.18$, $p < .001$), while the indirect effect through travel time is negative and significant ($\beta = -0.04$, $p < .01$).
As a result, the total effect is smaller but remains significant ($\beta = 0.14$, $p < .001$).
In addition, active mode use exerted a strong, positive direct effect on leisure participation ($\beta = 0.22$, $p < .001$), underscoring heterogeneity in activity outcomes beyond accessibility constraints. \par

\begin{table}[!ht]
  \centering
  \caption{Estimation results of SEM (significant at $\alpha = 0.05$).
  Standardized coefficients reported.}\label{tab:effects}
  \begin{tabular}{llll}
  \hline
  Effect on & Direct & Indirect & Total \\
  \hline
  Total travel time \\
  \quad STA value ($A_i$)          & $-0.33^{***}$ & --            & $-0.33^{***}$ \\
  \quad Active mode                & $-0.19^{***}$ & --            & $-0.19^{***}$ \\
  \quad Female                     & $-0.12^{***}$ & --            & $-0.12^{***}$ \\[6pt]
  Leisure participation \\
  \quad STA value ($A_i$)          & $0.18^{***}$  & $-0.04^{**}$  & $0.14^{***}$ \\
  \quad Total travel time          & $0.13^{***}$  & --            & $0.13^{***}$ \\
  \quad Active mode                & $0.22^{***}$  & --            & $0.22^{***}$ \\
  \quad Education: High            & $0.18^{***}$  & --            & $0.18^{***}$ \\
  \quad Poverty rate               & $-0.16^{***}$ & --            & $-0.16^{***}$ \\
  \quad Couple w/ children         & $-0.25^{***}$ & --            & $-0.25^{***}$ \\
  \hline
  \end{tabular}
  \begin{flushleft}
  \footnotesize Notes: STA = space--time accessibility. The indirect effect of STA operates through
  total travel time. Leisure participation is a latent construct indicated by location diversity and
  leisure duration.
  Significance levels: $^{*} p<0.05$, $^{**} p<0.01$, $^{***} p<0.001$.
  \end{flushleft}
\end{table}

While the total effect of STA on leisure participation was slightly offset by an indirect effect from shortened travel time, the decomposition highlights that STA does matter: it exerts a positive total effect on both leisure location diversity and activity duration ($\beta = 0.14$, $p < .001$).
These countervailing mechanisms suggest that accessibility shapes leisure activity participation in complex ways, rather than through a simple net effect.

\section{Conclusions}\label{sec:conclusion}
This study examined the impact of individual-based space–time accessibility (STA) on leisure activity participation, with a focus on the Paris region.
By leveraging high-resolution GPS data for 2,415 Paris residents, multimodal transport networks, and a capability-based accessibility framework, we demonstrated that STA---operationalized as the feasible set of leisure opportunities within individuals' time budgets---plays a critical role in shaping activity participation. \par

Our results offer robust empirical evidence that greater transport accessibility promotes more efficient travel (i.e., shorter travel times) to fulfill daily activity demand and supports broader participation in leisure activities (i.e., location diversity and activity duration).
Through structural equation modeling, we identified both direct and indirect effects: STA directly promotes a greater leisure participation, while it also reduces travel time, which is associated with lower participation.
However, the total effect of STA on leisure activity participation is significantly positive.
Such countervailing direct and indirect effects (via travel time) demonstrate that expanding opportunity sets does not linearly translate into broader participation, a finding that enriches the theoretical understanding of how capabilities convert into functionings through the mediating role of mobility behavior.
The observed spatial choices of leisure destinations further validated the STA construct as a realistic proxy for individuals' capability sets---particularly among active transport users, who showed a stronger alignment with modeled opportunity structures.
This alignment suggests that space–time prisms are not merely geometric constructs but behaviorally meaningful boundaries, providing rare empirical validation of the behavioral relevance of time-geographic constraints for discretionary activities. \par

Importantly, the study revealed substantial heterogeneity in leisure activity participation across population groups.
Education, local poverty, and household composition also directly shape leisure participation: positively for higher education and negatively for local poverty and couple-with-children households.
These reflect how human capital, economic deprivation, and caregiving responsibilities influence discretionary activity engagement beyond the indirect pathways through transport mode and accessibility.
These findings underscore the importance of considering both the spatial distribution of opportunities and the temporal feasibility of access when designing inclusive transport policies. \par

There are a few limitations in the present work.
First, our framework models accessibility and participation at the individual level and therefore cannot capture the joint, coordinated nature of many leisure activities — such as group outings or shared meals — where participation depends not only on one's own space–time constraints but also on those of companions, meaning that individual-level accessibility may be a necessary but insufficient condition for actual engagement.
Second, we acknowledge that the current STA operationalization captures only single-stop work–leisure–home chains on weekdays and does not account for weekend leisure patterns, multi-stop trip chaining, or leisure activities originating directly from home.
Third, non-workers are excluded from the analysis, despite potentially facing more severe accessibility constraints due to limited transport resources, reduced temporal flexibility, or geographic isolation from leisure opportunities.
Our findings on inclusive mobility, therefore, apply specifically to the commuting population; the mechanisms linking STA to leisure participation may operate differently for individuals whose daily routines are not structured around a work anchor.
Extending the STA framework to non-commuting populations would require alternative formulations, such as home-based prisms defined by discretionary time budgets and habitual activity anchors rather than fixed workplaces, representing an important direction for future research.

By embedding accessibility into a human capabilities framework, this work shifts the discourse from infrastructure provision toward understanding how individuals convert transport resources into meaningful participation.
Future research should extend this analysis to non-commuting populations, incorporate dynamic time budgets, and examine the longitudinal impacts of accessibility improvements on social inclusion and well-being.
As cities pursue more equitable and sustainable mobility systems, more nuanced accessibility metrics, such as space-time accessibility, can serve as valuable tools to identify gaps, monitor policy impacts, and prioritize interventions that expand real freedoms to engage in everyday urban life.

\section*{Data availability}
The data used in this study were provided through participation in the NetMob 2025 Data Challenge \citep{netmob25}, under a non-disclosure agreement (NDA) between all authors, Inria with IFPEN.
Due to licensing terms and privacy constraints governed by the European General Data Protection Regulation (GDPR), access to the data is restricted.
Venue locations and categories can be retrieved from Overture API.
Census data (income) were collected from Institut national de la statistique et des études économiques (INSEE) that is publicly available.
GTFS data were collected from transport.data.gouv.fr that are publicly available.
All data were utilized in accordance with the terms of service specified by their respective provider. \par

We adhered to the guidelines by the Chalmers Institutional Review Board (IRB) according to the Swedish Act (2003:460) concerning the ethical review of research involving humans, as well as the General Data Protection Regulation 2016/679 (GDPR).
According to the data applied, the study was exempt from ethical review under the Swedish Ethical Review Act (2003:460). \par

Python (version 3.11) code and R (version 4.5.1) code were used to analyse and visualize the data.
The accessibility-related travel times were calculated using r5r (version 2.3.0).
Code to reproduce our results is publicly available on \href{https://github.com/TheYuanLiao/netmob25}{GitHub Repository}.

\bibliographystyle{unsrtnat}
\bibliography{references}

@article{kitamura1984model,
  title={A model of daily time allocation to discretionary out-of-home activities and trips},
  author={Kitamura, Ryuichi},
  journal={Transportation Research Part B: Methodological},
  volume={18},
  number={3},
  pages={255--266},
  year={1984},
  publisher={Elsevier}
}

@article{ettema2007modelling,
  title={Modelling the joint choice of activity timing and duration},
  author={Ettema, Dick and Bastin, Fabian and Polak, John and Ashiru, Olu},
  journal={Transportation Research Part A: Policy and Practice},
  volume={41},
  number={9},
  pages={827--841},
  year={2007},
  publisher={Elsevier}
}

@article{schwanen2008fixed,
  title={How fixed is fixed? Gendered rigidity of space--time constraints and geographies of everyday activities},
  author={Schwanen, Tim and Kwan, Mei-Po and Ren, Fang},
  journal={Geoforum},
  volume={39},
  number={6},
  pages={2109--2121},
  year={2008},
  publisher={Elsevier}
}

@article{marchetti1994anthropological,
  title={Anthropological invariants in travel behavior},
  author={Marchetti, Cesare},
  journal={Technological forecasting and social change},
  volume={47},
  number={1},
  pages={75--88},
  year={1994},
  publisher={Elsevier}
}

@article{schafer2000future,
  title={The future mobility of the world population},
  author={Schafer, Andreas and Victor, David G},
  journal={Transportation research part a: policy and practice},
  volume={34},
  number={3},
  pages={171--205},
  year={2000},
  publisher={Elsevier}
}

@book{hair2010multivariate,
    title={Multivariate Data Analysis},
    author={Hair, Joseph F. and Black, William C. and Babin, Barry J. and Anderson, Rolph E.},
    year={2010},
    edition={7th},
    publisher={Pearson Prentice Hall},
    address={Upper Saddle River, NJ}
  }

@book{ben1985discrete,
  title={Discrete choice analysis: theory and application to travel demand},
  author={Ben-Akiva, Moshe E and Lerman, Steven R},
  volume={9},
  year={1985},
  publisher={MIT press}
}

@misc{INSEE_RevenusPauvreteIris2021,
  title        = {Revenus, pauvret{\'e} et niveau de vie en 2021 (Iris)},
  author       = {{Institut national de la statistique et des \'{e}tudes \'{e}conomiques (INSEE)}},
  year         = {2024},
  howpublished = {Web page},
  url          = {https://www.insee.fr/fr/statistiques/8229323},
}

@article{stauffacher2005diversity,
  title={The diversity of travel behaviour: motives and social interactions in leisure time activities},
  author={Stauffacher, Michael and Schlich, Robert and Axhausen, Kay W and Scholz, Roland W},
  journal={Arbeitsberichte Verkehrs-und Raumplanung},
  volume={328},
  year={2005},
  publisher={ETH, Eidgen{\"o}ssische Technische Hochschule Z{\"u}rich, IVT, Institut f{\"u}r~…}
}

@article{de2016travel,
  title={Travel mode choice and travel satisfaction: bridging the gap between decision utility and experienced utility},
  author={De Vos, Jonas and Mokhtarian, Patricia L and Schwanen, Tim and Van Acker, Veronique and Witlox, Frank},
  journal={Transportation},
  volume={43},
  number={5},
  pages={771--796},
  year={2016},
  publisher={Springer}
}

@article{hu1999cutoff,
  title={Cutoff criteria for fit indexes in covariance structure analysis: Conventional criteria versus new alternatives},
  author={Hu, Li-tze and Bentler, Peter M},
  journal={Structural equation modeling: a multidisciplinary journal},
  volume={6},
  number={1},
  pages={1--55},
  year={1999},
  publisher={Taylor \& Francis}
}

@article{Azmoodeh_et_al_2023,
  author = {Azmoodeh, Mohammad and Haghighi, Farshidreza and Motieyan, Hamid},
  year = {2023},
  title = {The capability approach and social equity in transport: Understanding factors affecting capabilities of urban residents, using structural equation modeling},
  journal = {Transport Policy},
  volume = {142},
  pages = {137--151},
  doi = {10.1016/j.tranpol.2023.08.010},
  publisher = {Elsevier Ltd}
}

@article{Golob_2003,
  author = {Golob, Thomas F.},
  year = {2003},
  title = {Structural equation modeling for travel behavior research},
  journal = {Transportation Research Part B: Methodological},
  volume = {37},
  pages = {1--25},
  doi = {10.1016/S0191-2615(01)00046-7},
  publisher = {Elsevier}
}

@article{Kroesen_Van_Wee_2022,
  author = {Kroesen, Maarten and Van Wee, Bert},
  year = {2022},
  title = {Understanding how accessibility influences health via active travel: Results from a structural equation model},
  journal = {Journal of Transport Geography},
  volume = {102},
  pages = {103379},
  doi = {10.1016/j.jtrangeo.2022.103379},
  publisher = {Elsevier Ltd}
}

@article{Li_et_al_2024,
  author = {Li, Linna and Cai, Jiayuan and Chen, Wenfeng},
  year = {2024},
  title = {How does transport development contribute to rural income in China? Evidence from county-level analysis using structural equation model},
  journal = {Travel Behaviour and Society},
  volume = {34},
  pages = {100708},
  doi = {10.1016/j.tbs.2023.100708},
  publisher = {Elsevier Ltd}
}

@article{Song_et_al_2016,
  author = {Song, Siqi and Diao, Mi and Feng, Chen-Chieh},
  year = {2016},
  title = {Individual transport emissions and the built environment: A structural equation modelling approach},
  journal = {Transportation Research Part A: Policy and Practice},
  volume = {92},
  pages = {206--219},
  doi = {10.1016/j.tra.2016.08.005},
  publisher = {Elsevier Ltd}
}

@article{Zhang_et_al_2025,
  author = {Zhang, Yingheng and Li, Haojie and Ren, Gang},
  year = {2025},
  title = {Ex-post evaluation of transport interventions with causal mediation analysis},
  journal = {Transportation},
  volume = {52},
  pages = {93--126},
  doi = {10.1007/s11116-023-10413-0},
  publisher = {Springer Science+Business Media, LLC, part of Springer Nature}
}

@article{kwan1998space,
  title={Space-time and integral measures of individual accessibility: a comparative analysis using a point-based framework},
  author={Kwan, Mei-Po},
  journal={Geographical analysis},
  volume={30},
  number={3},
  pages={191--216},
  year={1998},
  publisher={Wiley Online Library}
}

@article{mccormack2011search,
  title={In search of causality: a systematic review of the relationship between the built environment and physical activity among adults},
  author={McCormack, Gavin R and Shiell, Alan},
  journal={International journal of behavioral nutrition and physical activity},
  volume={8},
  number={1},
  pages={125},
  year={2011},
  publisher={Springer}
}

@article{zhang2022eliminating,
  title={Eliminating barriers to nighttime activity participation: the case of on-demand transit in Belleville, Canada},
  author={Zhang, Yixue and Farber, Steven and Young, Mischa},
  journal={Transportation},
  volume={49},
  number={5},
  pages={1385--1408},
  year={2022},
  publisher={Springer}
}

@article{allen2020planning,
  title={Planning transport for social inclusion: An accessibility-activity participation approach},
  author={Allen, Jeff and Farber, Steven},
  journal={Transportation Research Part D: Transport and Environment},
  volume={78},
  pages={102212},
  year={2020},
  publisher={Elsevier}
}

@article{luz2022does,
  title={Does better accessibility help to reduce social exclusion? Evidence from the city of S{\~a}o Paulo, Brazil},
  author={Luz, Greg{\'o}rio and Barboza, Matheus HC and Portugal, Licinio and Giannotti, Mariana and Van Wee, Bert},
  journal={Transportation research part A: policy and practice},
  volume={166},
  pages={186--217},
  year={2022},
  publisher={Elsevier}
}

@article{miller1991modelling,
  title={Modelling accessibility using space-time prism concepts within geographical information systems},
  author={Miller, Harvey J},
  journal={International Journal of Geographical Information System},
  volume={5},
  number={3},
  pages={287--301},
  year={1991},
  publisher={Taylor \& Francis}
}

@article{lee2022third,
  title={Third place and psychological well-being: The psychological benefits of eating and drinking places for university students in Southern California, USA},
  author={Lee, Narae},
  journal={Cities},
  volume={131},
  pages={104049},
  year={2022},
  publisher={Elsevier}
}

@article{nilforoshan2023human,
  title={Human mobility networks reveal increased segregation in large cities},
  author={Nilforoshan, Hamed and Looi, Wenli and Pierson, Emma and Villanueva, Blanca and Fishman, Nic and Chen, Yiling and Sholar, John and Redbird, Beth and Grusky, David and Leskovec, Jure},
  journal={Nature},
  volume={624},
  number={7992},
  pages={586--592},
  year={2023},
  publisher={Nature Publishing Group UK London}
}

@article{liao2025socio,
  title={Socio-spatial segregation and human mobility: A review of empirical evidence},
  author={Liao, Yuan and Gil, Jorge and Yeh, Sonia and Pereira, Rafael HM and Alessandretti, Laura},
  journal={Computers, Environment and Urban Systems},
  volume={117},
  pages={102250},
  year={2025},
  publisher={Elsevier}
}

@article{luz2022understanding,
  title={Understanding transport-related social exclusion through the lens of capabilities approach},
  author={Luz, Gregorio and Portugal, Licinio},
  journal={Transport Reviews},
  volume={42},
  number={4},
  pages={503--525},
  year={2022},
  publisher={Taylor \& Francis}
}

@article{kwan1999gender,
  title={Gender and individual access to urban opportunities: a study using space--time measures},
  author={Kwan, Mei-Po},
  journal={The Professional Geographer},
  volume={51},
  number={2},
  pages={210--227},
  year={1999},
  publisher={Wiley Online Library}
}

@article{gallego2023social,
  title={Social Inequality in Popular Neighborhoods: A Pre-and Post-Pandemic Perspective from Joint Accessibility},
  author={Gallego M{\'e}ndez, Jorge and Garc{\'\i}a-Moreno, Lina M and Murillo-Hoyos, Jackeline and Jaramillo Molina, Ciro},
  journal={Sustainability},
  volume={15},
  number={13},
  pages={10587},
  year={2023},
  publisher={MDPI}
}

@article{pereira2017distributive,
  title={Distributive justice and equity in transportation},
  author={Pereira, Rafael HM and Schwanen, Tim and Banister, David},
  journal={Transport reviews},
  volume={37},
  number={2},
  pages={170--191},
  year={2017},
  publisher={Taylor \& Francis}
}

@article{ryan2023accessibility,
  title={Accessibility and space-time differences in when and how different groups (choose to) travel},
  author={Ryan, Jean and Pereira, Rafael HM and Andersson, Magnus},
  journal={Journal of Transport Geography},
  volume={111},
  pages={103665},
  year={2023},
  publisher={Elsevier}
}

@article{botta2021modelling,
  title={Modelling urban vibrancy with mobile phone and OpenStreetMap data},
  author={Botta, Federico and Guti{\'e}rrez-Roig, Mario},
  journal={Plos one},
  volume={16},
  number={6},
  pages={e0252015},
  year={2021},
  publisher={Public Library of Science San Francisco, CA USA}
}

@article{saraiva2022accessibility,
  title={Accessibility in S{\~a}o Paulo: an individual road to equity?},
  author={Saraiva, Marcus and Barros, Joana},
  journal={Applied geography},
  volume={144},
  pages={102731},
  year={2022},
  publisher={Elsevier}
}

@article{mcguckin1999examining,
  title={Examining trip-chaining behavior: Comparison of travel by men and women},
  author={McGuckin, Nancy and Murakami, Elaine},
  journal={Transportation research record},
  volume={1693},
  number={1},
  pages={79--85},
  year={1999},
  publisher={SAGE Publications Sage CA: Los Angeles, CA}
}

@article{oldenburgThirdPlace1982,
  title = {The Third Place},
  author = {Oldenburg, Ramon and Brissett, Dennis},
  year = {1982},
  journal = {Qualitative Sociology},
  volume = {5},
  number = {4},
  pages = {265--284},
  issn = {1573-7837},
  doi = {10.1007/BF00986754},
  urldate = {2025-03-11},
  langid = {english}
}

@article{reimers2014proximity,
  title={Proximity to sports facilities and sports participation for adolescents in Germany},
  author={Reimers, Anne K and Wagner, Matthias and Alvanides, Seraphim and Steinmayr, Andreas and Reiner, Miriam and Schmidt, Steffen and Woll, Alexander},
  journal={PLoS One},
  volume={9},
  number={3},
  pages={e93059},
  year={2014},
  publisher={Public Library of Science San Francisco, USA}
}

@article{joly2016intensive,
  title={Intensive travel time: an obligation or a choice?},
  author={Joly, Iraga{\"e}l and Vincent-Geslin, St{\'e}phanie},
  journal={European Transport Research Review},
  volume={8},
  number={1},
  pages={10},
  year={2016},
  publisher={Springer}
}

@article{althoff2025countrywide,
  title={Countrywide natural experiment links built environment to physical activity},
  author={Althoff, Tim and Ivanovic, Boris and King, Abby C and Hicks, Jennifer L and Delp, Scott L and Leskovec, Jure},
  journal={Nature},
  pages={1--7},
  year={2025},
  publisher={Nature Publishing Group UK London}
}

@misc{insee_pop_2020_regions,
  author       = {{INSEE}},
  title        = {Populations légales des régions en 2020 – Recensement de la population},
  year         = {2022},
  howpublished = {\url{https://www.insee.fr/fr/statistiques/6683011?sommaire=6683037}},
  note         = {Accessed: 2025-08-19}
}

@misc{idfmmob_mobility_as_a_service_2023,
  author       = {{Île-de-France Mobilités}},
  title        = {{Reference Guide for Mobility-as-a-Service (MaaS)}},
  year         = {2023},
  month        = feb,
  howpublished = {PDF document on the PRIM platform},
  note         = {Online at: \url{https://prim.iledefrance-mobilites.fr/content/files/2023/02/IDFM_Reference-guide-for-Mobility-as-a-Service_english.pdf}, accessed 2025-08-19},
}

@article{yin2023multimodal,
  title={What are the multimodal patterns of individual mobility at the day level in the Paris region? A two-stage data-driven approach based on the 2018 Household Travel Survey},
  author={Yin, Biao and Leurent, Fabien},
  journal={Transportation},
  volume={50},
  number={4},
  pages={1497--1526},
  year={2023},
  publisher={Springer}
}

@misc{lagrandeconversation2023,
  author       = {{La Grande Conversation}},
  title        = {Fewer Parisians but More Greater Parisians: Density in the Île-de-France},
  year         = {2023},
  howpublished = {\url{https://www.lagrandeconversation.com/en/society/fewer-parisians-but-more-greater-parisians-density-in-the-ile-de-france}},
  note         = {Accessed: 2025-08-18}
}

@misc{uberh3,
  author       = {{Uber Technologies, Inc.}},
  title        = {H3: A Hexagonal Hierarchical Spatial Index},
  year         = {2018},
  howpublished = {\url{https://h3geo.org/}},
  note         = {Accessed: 2025-08-18}
}

@misc{overturemaps,
  author       = {{Overture Maps Foundation}},
  title        = {Overture Maps API},
  year         = {2023},
  howpublished = {\url{https://overturemaps.org/}},
  note         = {Accessed: 2025-08-18}
}

@misc{netmob25,
title={The NetMob25 Dataset: A High-resolution Multi-layered View of Individual Mobility in Greater Paris Region},
author={Chasse, Alexandre and Kouam, Anne J. and Viana, Aline C. and Stanica, Razvan and Lobato, Wellington V. and Ramos, Geymerson and Deperle, Geoffrey and Bouroudi, Abdelmounaim and Bussod, Suzanne and Molano, Fernando},
year={2025},
eprint={2506.05903},
archivePrefix={arXiv},
primaryClass={cs.CY},
url={https://arxiv.org/abs/2506.05903}
}

@article{r5r,
title = {r5r: Rapid Realistic Routing on Multimodal Transport
  Networks with R5 in R},
author = {Rafael H. M. Pereira and Marcus Saraiva and Daniel
  Herszenhut and Carlos Kaue Vieira Braga and Matthew Wigginton
  Conway},
journal = {Findings},
year = {2021},
doi = {10.32866/001c.21262},
url = {https://doi.org/10.32866/001c.21262},
}

@article{levinson2020transport,
  title={Transport access manual: A guide for measuring connection between people and places},
  author={Levinson, David and King, David},
  year={2020},
  publisher={Committee of the Transport Access Manual, University of Sydney}
}

@article{victoriano2020time,
  title={Time, space, money, and social interaction: Using machine learning to classify people’s mobility strategies through four key dimensions},
  author={Victoriano, Rodrigo and Paez, Antonio and Carrasco, Juan-Antonio},
  journal={Travel Behaviour and Society},
  volume={20},
  pages={1--11},
  year={2020},
  publisher={Elsevier}
}

@article{kim2003space,
  title={Space-time accessibility measures: A geocomputational algorithm with a focus on the feasible opportunity set and possible activity duration},
  author={Kim, Hyun-Mi and Kwan, Mei-Po},
  journal={Journal of geographical Systems},
  volume={5},
  number={1},
  pages={71--91},
  year={2003},
  publisher={Springer}
}

@article{fransen2018spatio,
  title={A spatio-temporal accessibility measure for modelling activity participation in discretionary activities},
  author={Fransen, Koos and Farber, Steven and Deruyter, Greta and De Maeyer, Philippe},
  journal={Travel behaviour and society},
  volume={10},
  pages={10--20},
  year={2018},
  publisher={Elsevier}
}

@article{neutens2010equity,
  title={Equity of urban service delivery: a comparison of different accessibility measures},
  author={Neutens, Tijs and Schwanen, Tim and Witlox, Frank and De Maeyer, Philippe},
  journal={Environment and Planning a},
  volume={42},
  number={7},
  pages={1613--1635},
  year={2010},
  publisher={SAGE Publications Sage UK: London, England}
}

@article{barboza2024comparative,
  title={A comparative analysis of leisure accessibility and equity impacts using location-based and space--time accessibility metrics},
  author={Barboza, Matheus HC and Giannotti, Mariana and Grigolon, Anna B and Geurs, Karst T},
  journal={Transportation Research Part A: Policy and Practice},
  volume={190},
  pages={104237},
  year={2024},
  publisher={Elsevier}
}

@article{tomasiello2023unfolding,
  title={Unfolding time, race and class inequalities to access leisure},
  author={Tomasiello, Diego Bogado and Giannotti, Mariana},
  journal={Environment and Planning B: Urban Analytics and City Science},
  volume={50},
  number={4},
  pages={927--941},
  year={2023},
  publisher={SAGE Publications Sage UK: London, England}
}

@article{chen2021effects,
  title={Effects of built environment on activity participation under different space-time constraints: A case study of Guangzhou, China},
  author={Chen, Zifeng and Yeh, Anthony Gar-On},
  journal={Travel Behaviour and Society},
  volume={22},
  pages={84--93},
  year={2021},
  publisher={Elsevier}
}

@article{farber2013social,
  title={The social interaction potential of metropolitan regions: A time-geographic measurement approach using joint accessibility},
  author={Farber, Steven and Neutens, Tijs and Miller, Harvey J and Li, Xiao},
  journal={Annals of the Association of American Geographers},
  volume={103},
  number={3},
  pages={483--504},
  year={2013},
  publisher={Taylor \& Francis}
}

@article{van2022disentangling,
  title={Disentangling barrier effects of transport infrastructure: synthesising research for the practice of impact assessment},
  author={van Eldijk, Job and Gil, Jorge and Marcus, Lars},
  journal={European transport research review},
  volume={14},
  number={1},
  pages={1},
  year={2022},
  publisher={Springer}
}

@book{huntington2021effect,
  title={The effect: An introduction to research design and causality},
  author={Huntington-Klein, Nick},
  year={2021},
  publisher={Chapman and Hall/CRC}
}

\section*{Acknowledgements}
This research is funded by the Swedish Research Council (Project Number 2022-06215).

\section*{Author contributions}
Y.L. conceptualized the study.
All authors designed the methods.
Y.L. processed the data and the model.
All authors wrote the manuscript.

\section*{Competing interests}
The authors declare that there are no conflicts of interest.

\section*{Additional information}
Correspondence and requests for materials should be addressed to Y.L.

\appendix
\renewcommand{\thefigure}{A.\arabic{figure}}
\renewcommand{\thetable}{A.\arabic{table}}
\setcounter{figure}{0}
\setcounter{table}{0}
\section{Sensitivity test results}\label{seca:sta}
\subsection{Space-time accessibility distributions}
\begin{figure}[!htp]
\centering
\includegraphics[width=1\linewidth]{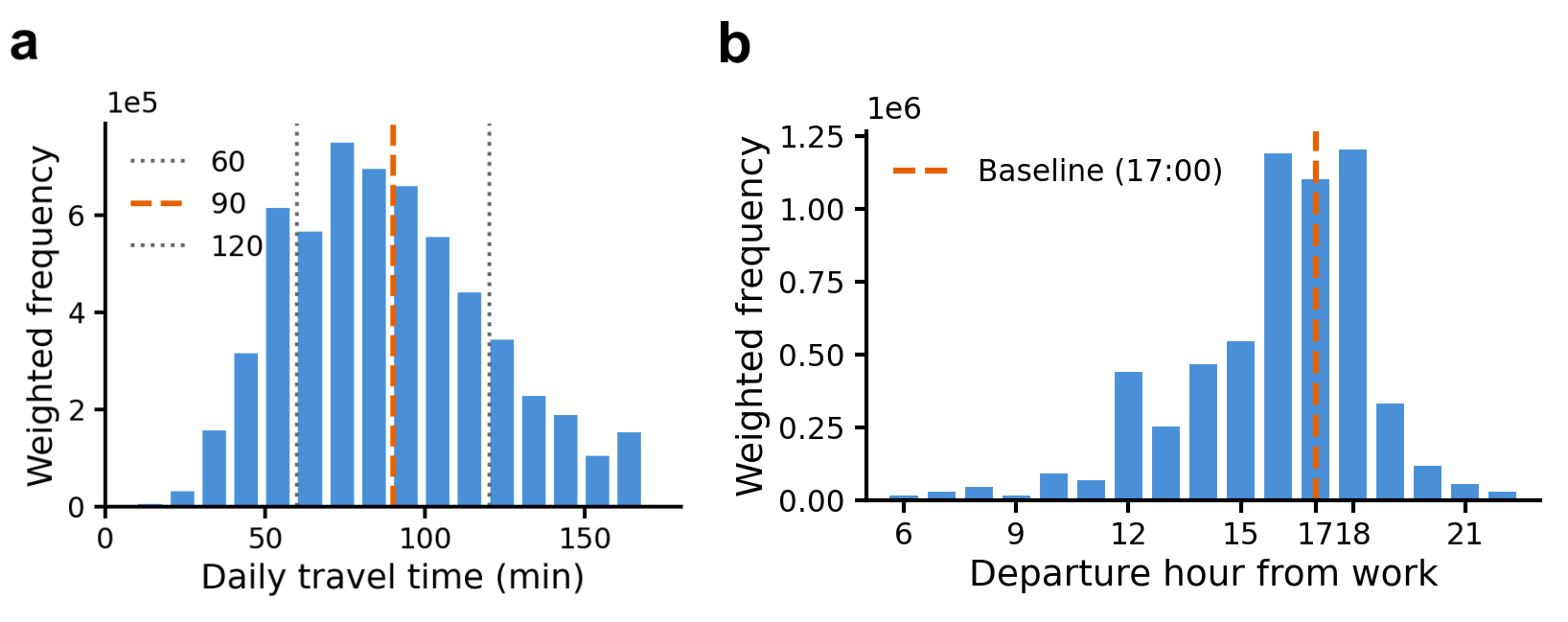}
\caption{\textbf{Parameter choices of space–time accessibility.}
\textbf{a}, Distribution of travel time budget.
\textbf{b}, Distribution of departure time from work.}\label{fig:sta_paras}
\end{figure}

To assess the robustness of STA to its two main parametric assumptions, we conducted sensitivity analyses varying departure time and time budget (Figure \ref{fig:sta_paras}). \par

First, varying the time budget from 60 to 120 minutes at a fixed 17:00 departure, the share of individuals with STA > 0 increases from 28.7\% to 59.7\% for car users and from 13.1\% to 49.4\% for public transit users.
The baseline of 90 minutes, approximating the empirical median of observed daily travel times (82 minutes), thus produces intermediate coverage rather than ceiling or floor effects.
Second, varying the departure hour from 16:00 to 18:00 for public transit users at a fixed 90-minute budget, both the mean number of accessible opportunities (979-1,029) and the share with non-zero STA (32\%) remain stable, indicating that the measure reflects underlying spatial structure rather than schedule-specific artifacts.
Together, these results confirm that while absolute STA magnitudes scale with the time budget as expected, the qualitative patterns are robust to reasonable parameter variation.

\begin{figure}[!htp]
\centering
\includegraphics[width=1\linewidth]{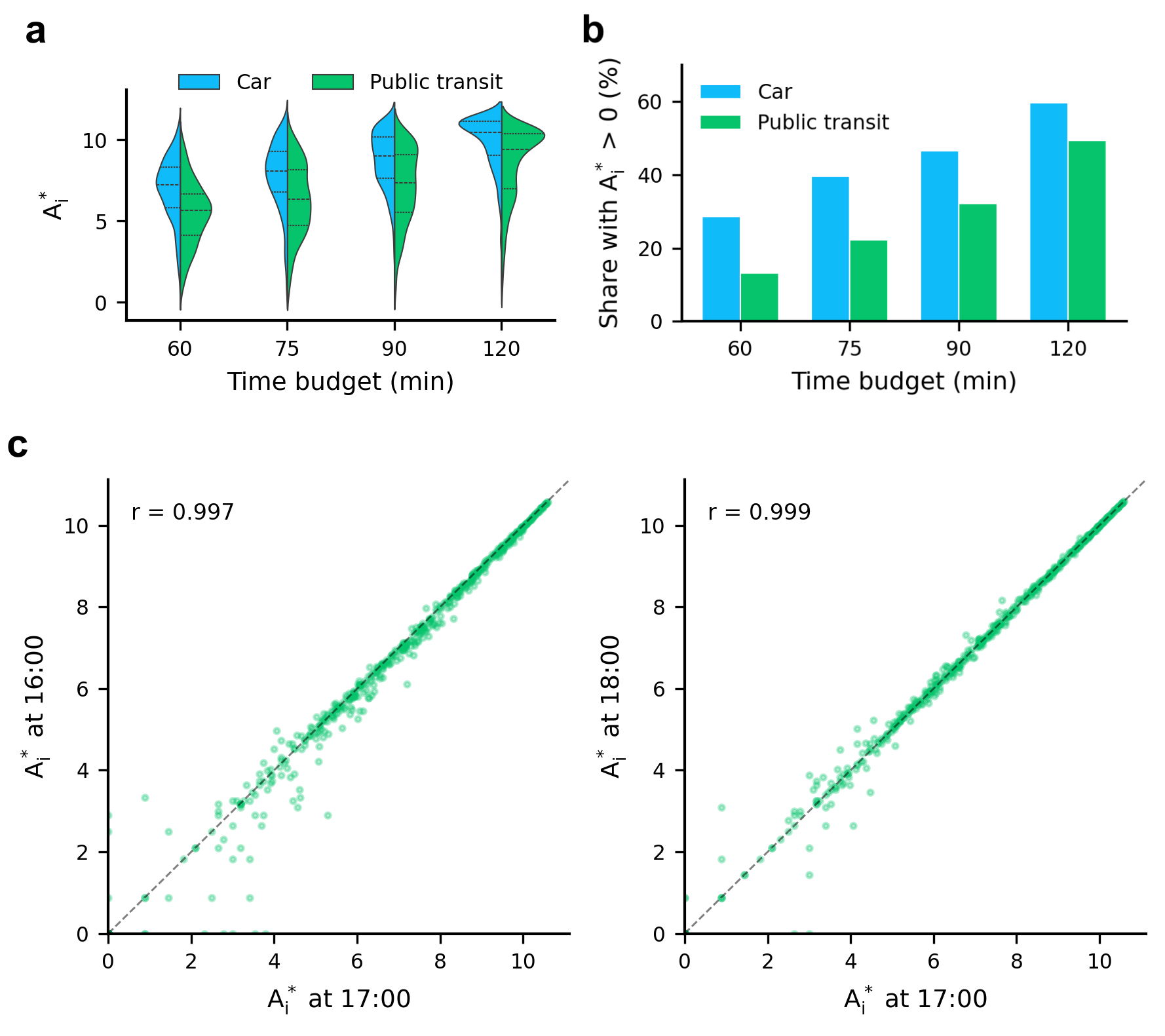}
\caption{\textbf{Sensitivity of space–time accessibility to parameter choices.}
\textbf{a}, Distribution of STA (IHS-transformed) by travel time budget, for individuals with $A_i^*> 0$.
Split violins compare car (blue) and public transit (green) users.
\textbf{b}, Share of individuals with $A_i^* > 0$ by travel time budget and mode.
\textbf{c}, Pairwise comparison of STA at different departure hours (16:00, 17:00, 18:00) for public transit users with 90-minute budget.
Diagonal line indicates perfect agreement; correlation coefficients shown.}\label{fig:sensitivity}
\end{figure}

\subsection{SEM outcomes}\label{seca:sem}
To assess the robustness of our findings to the operationalization of space–time accessibility, we conducted sensitivity analyses varying the travel time budget parameter.
Given the limited impact of departure time (16:00--18:00, see Figure \ref{fig:sensitivity}), we re-estimated the full SEM specification under two alternative scenarios fixing departure time as 17:00 in the baseline: a restrictive 60-minute budget representing time-constrained individuals, and a generous 120-minute budget representing those with greater temporal flexibility.
For each specification, we computed STA using the same methodology and re-estimated the latent variable SEM with identical model structure, covariates, and estimation procedure. \par

Model fit remained excellent across all time budget specifications (Table \ref{tab:sem_fit_sensitivity}), indicating that the model structure is robust to the choice of time budget parameter. \par

  \begin{table}[!ht]
  \centering
  \caption{SEM model fit across time budget specifications.
  Departure time is 17:00.
  CFI = Comparative Fit Index; TLI = Tucker-Lewis Index; RMSEA = Root Mean Square Error of Approximation; SRMR = Standardized Root Mean Square Residual.
  All models use DWLS estimation with sampling weights.
  $N = 2{,}415$.}\label{tab:sem_fit_sensitivity}
  \begin{tabular}{lccc}
  \hline
  Budget (min) & 60 & 90 (baseline) & 120 \\
  \hline
  CFI   & .988 & .987 & .986 \\
  TLI   & .958 & .954 & .951 \\
  RMSEA & .027 & .029 & .030 \\
  \quad 90\% CI & [.015, .039] & [.018, .041] & [.019, .042] \\
  SRMR  & .013 & .014 & .014 \\
  \hline
  \end{tabular}
  \end{table}

The pattern of effects are consistent across specifications (Table \ref{tab:sem_effects_sensitivity}).
All three models show a significant positive direct effect of STA on leisure participation and a significant negative indirect effect operating through travel time, yielding a positive total effect.
The countervailing mechanism—whereby higher accessibility directly increases leisure engagement but indirectly reduces it by lowering travel time—is replicated under all budget assumptions. \par

  \begin{table}[!ht]
  \centering
  \caption{Effect of space--time accessibility on leisure participation across time budget specifications.
  Standardized coefficients reported for STA ($A_i^*$) effect on latent leisure participation.
  Direct effect: STA $\rightarrow$ Leisure participation.
  Indirect effect: STA $\rightarrow$ Total travel time $\rightarrow$ Leisure participation.
  Significance levels: $^{*}p<0.05$, $^{**}p<0.01$, $^{***}p<0.001$.}\label{tab:sem_effects_sensitivity}
  \begin{tabular}{lccc}
  \hline
  Effect & 60 min & 90 min (baseline) & 120 min \\
  \hline
  Direct
    & $.125^{***}$ & $.181^{***}$ & $.165^{***}$ \\
  Indirect
    & $-.027^{**}$ & $-.043^{**}$ & $-.042^{**}$ \\
  Total
    & $.098^{***}$ & $.138^{***}$ & $.123^{***}$ \\
  \hline
  \end{tabular}
  \end{table}

\begin{figure*}[!htp]
\centering
\includegraphics[width=1\linewidth]{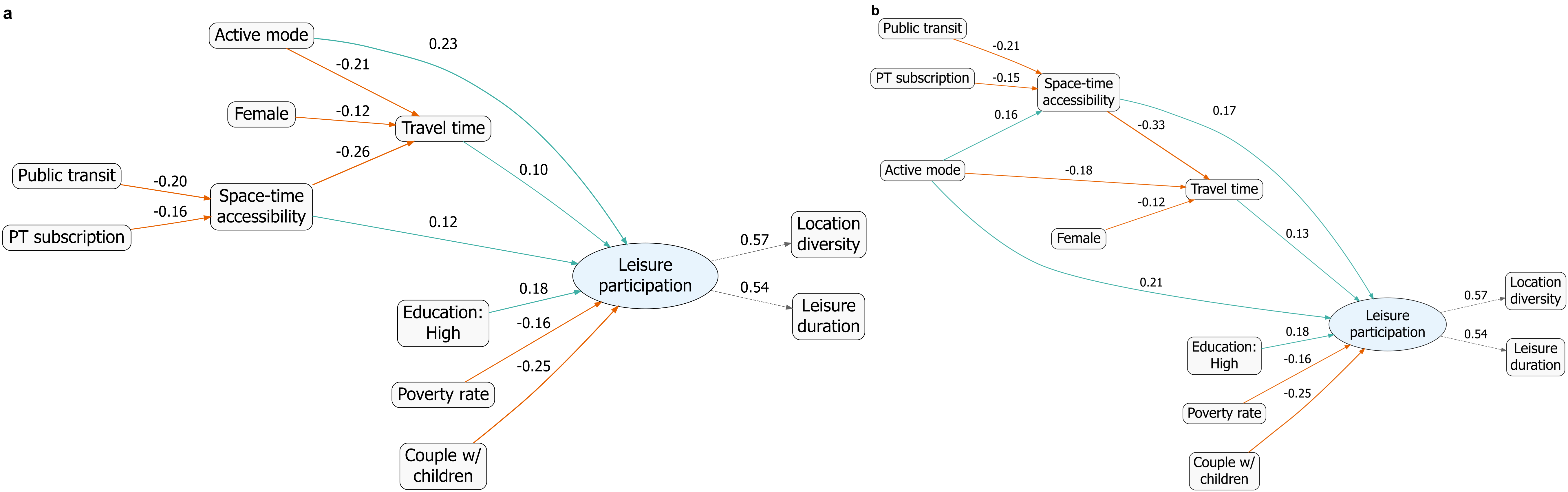}
\caption{\textbf{Path diagrams for SEM sensitivity analysis under alternative time budget specifications.}
  \textbf{a}, Restrictive 60-minute travel time budget.
  \textbf{b}, Generous 120-minute travel time budget.
  Standardized coefficients shown for paths with $p < 0.05$ and $|\beta| > 0.10$.
  Teal arrows indicate positive effects; orange arrows indicate negative effects.
  Arrow thickness proportional to effect magnitude.
  Dashed lines denote measurement model loadings.
  The latent leisure participation construct is indicated by location diversity and leisure duration.}\label{fig:sem_sensitivity}
\end{figure*}

Effect magnitudes vary with the time budget but with stable direction and strengths (Figure \ref{fig:sem_sensitivity}).
These findings support the 90-minute threshold as an appropriate balance between capturing meaningful accessibility variation and avoiding floor or ceiling effects, while confirming that the substantive conclusions are robust to reasonable alternative specifications.

\end{document}